\documentclass[aps,pre,twocolumn,floatfix,superscriptaddress,showpacs,showkeys]{revtex4-1}
\usepackage{epsf,amsmath,amssymb,verbatim,color,multirow,pifont,graphicx,cleveref,tabularx,cancel,soul}

\crefname{equation}{Eq.}{Eqs.}
\crefname{section}{Sec.}{Secs.}
\crefname{figure}{Fig.}{Figs.}
\crefname{table}{Table}{Tables.}

\newcommand{\ro}{\rho_0}
\newcommand{\ps}{P_{\rm S}}
\newcommand{\pw}{P_{\rm W}}
\newcommand{\pr}{P_{\rm R}}
\newcommand{\pinf}{P_{\infty}}
\newcommand{\pinfn}{P_{\infty,N}}

\newcommand{\nubar}{\bar \nu}
\newcommand{\self}{self-consistency }

\begin{document}

\title{Mixed-order phase transition in a two-step contagion model with single infectious seed}
\author{Wonjun Choi}
\affiliation{CCSS,  CTP and  Department of Physics and Astronomy, Seoul National University, Seoul 08826, Korea}
\author{Deokjae Lee}
\affiliation{CCSS,  CTP and Department of Physics and Astronomy, Seoul National University, Seoul 08826, Korea}
\author{B. Kahng}
\email{bkahng@snu.ac.kr}
\affiliation{CCSS,  CTP and  Department of Physics and Astronomy, Seoul National University, Seoul 08826, Korea}
\date{\today}

\begin{abstract}
A hybrid phase transition (HPT) that exhibits properties of continuous and discontinuous phase transitions at the same transition point has been observed in diverse complex systems. Previous studies of the HPTs  on complex networks mainly focused on whether the order parameter is continuous or discontinuous. However, more careful and fundamental questions on the critical behaviors of the HPT  such as how the divergences of the susceptibility and of the correlation size are affected by the discontinuity of the order parameter have been addressed.  Here, we consider a generalized epidemic model that is known to exhibit a discontinuous transition as a spinodal transition. Performing extensive numerical simulations and using finite-size scaling analysis, we examine diverging behaviors of the susceptibility and the correlation size. We find that when there is one infectious node and under a certain condition, the order parameter can exhibit a discontinuous jump but does not exhibit any critical behavior before or after the jump. This feature differs from what we observed in HPTs  in the percolation pruning process. However, critical behavior appears in the form of a power-law behavior of the outbreak size distribution. The mean outbreak size, corresponding to the susceptibility, diverge following the conventional percolation behavior. Thus a mixed-order transition occurs. The hyperscaling relation does not hold. 
\end{abstract}


\maketitle

\section{Introduction}
Hybrid phase transitions (HPTs) have been observed in diverse models on complex networks, for instance, $k$-core percolation~\cite{kcore1,kcore2,kcore_prx}, the cascading failure (CF) model~\cite{buldyrev,grassberger,mcc_dj} on interdependent networks and a synchronization model~\cite{moreno}. Those models have provided some basic idea to understand drastic changing phenomena in real-world complex systems such as jamming transitions~\cite{jamming1,jamming2}, blackout of power-grid systems~\cite{buldyrev} and so on. In the above two percolation models~\cite{kcore2,buldyrev}, as nodes or links are removed one by one from a certain control parameter value $r$ above a transition point $r_c$, the order parameter decreases continuously, but it also displays discontinuous feature at $r_c$. Thus,  the order parameter $m(r)$ behaves as follows: 
\begin{equation}
m(r)=\left\{
\begin{array}{lr}
m_0+b(r-r_c)^{\beta_m} & ~{\rm for}~~ r \ge r_c, \\
0 & ~{\rm for}~~  r < r_c,  
\end{array}
\right.
\label{order}
\end{equation} 
where $m_0$ and $b$ are constants, and $\beta_m$ is the critical exponent of the order parameter. The hybrid behavior of the order parameter was mainly issued in early researches. However, recently divergent behaviors of the fluctuations of the order parameter and of the correlation size (i.e., the number of nodes correlated) at the transition point and scaling relations have also been focused~\cite{mcc_dj}. The critical exponents associated with those quantities $\{\gamma_m, \nubar_m\}$, respectively were defined. On the other hand, when a node is deleted, it can trigger cascading failures. In the critical region, the avalanche size distribution follows a power law as $p_s\sim s^{-\tau_a}$ with $\tau_a=3/2$; however, the cluster size distribution does not exhibit power-law behavior. This feature is unconventional from the perspective of the percolation theory for the ordinary percolation~\cite{perc_review}. The power-law behavior in the avalanche dynamics yields another type of critical behavior, which requires  another set of the exponents $\{\tau_a, \sigma_a, \gamma_a, \nubar_a\}$~\cite{mcc_dj}. It reveals that for the CF model, the critical exponents in the set for the order parameter satisfy scaling relations by themselves but those in the set of the avalanche dynamics do not satisfy hyperscaling relations. Moreover, the two sets of the exponents are not completely independent, but they are coupled as $1-\beta_m=\gamma_a$.   

It was proposed that a discontinuous percolation transition cannot occur when its occupation rule is local~\cite{riordan}. However, when more than one species of particles cooperatively occupy each node, a discontinuous percolation transition can occur even though the dynamic rule is local~\cite{two_type, experiment}. Along these lines, the conventional susceptible-infected-susceptible (SIR) model, which exhibits a continuous percolation transition, was generalized in several forms to produce a discontinuous phase transition or HPT~\cite{janssen,dodds}. One model introduced in \cite{janssen} (called the SWIR  model) contains an  intermediate state called the weakened state (symbolized as W) between susceptible state (S) and infectious state (I). A node in state W becomes more easily infected than others in state S, so the reaction changes rapidly to the recovered state R, leading to a discontinuous transition~\cite{janssen,hasegawa,janssen_spinodal,chung}. Another model introduced in Ref.~\cite{grassberger_epl_2013} allows the occupations of two pathogens I$_{\rm a}$ and  I$_{\rm_b}$ on each node instead of single pathogen I.  
When a node is in the two-pathogen state, it can more easily change its state to state R, which leads to a discontinuous transition. 

Recent studies of both of these generalized contagion models focused on the behavior of the order parameter using the local tree approximation. However, to understand the critical behavior of the HPT thoroughly, one needs to check whether other physical quantities such as the susceptibility and the correlation size conform to the conventional critical properties and thus whether their exponents satisfy the scaling relations or not. To check the criticality, here we investigate the behavior of the order parameter and the scaling relations of the critical exponents for the SWIR  model~\cite{janssen}. We find that the order parameter does not follow the formula (\ref{order}); further, the fluctuation of the order parameter does not diverge.  Thus the order parameter does not exhibit the feature of the HPT. However, the probability $\pinf(r)$ that a macroscopic-scale outbreak (called an infinite outbreak hereafter) occurs as a function of the infection probability $r$ exhibits a critical behavior~\cite{grassberger_nphy}. An infinite outbreak is required for the order parameter to jump from zero to a finite value. It was shown~\cite{grassberger_nphy} that the probability $\pinf(r)$ is nothing but the spanning probability of the percolation. Thus, the critical exponents of the ordinary percolation $\{\beta_p, \gamma_p, \nubar_p\}$ govern the critical behavior of the SWIR model. 

From a single source of epidemic spreading, finite outbreaks can occur and their sizes (corresponding to finite avalanche sizes in the CF model) are heterogeneous according to a power law $p_s(r)\sim s^{-\tau_a}\exp(-s/s^*)$, where $\tau_a=3/2$ and $s^*\sim (r-r_c)^{-1/\sigma_a}$ with $r_c$ being a transition point. Using finite size scaling analysis, we can obtain the critical exponents $\{\gamma_a, \nubar_a\}$. Thus, the two sets of critical exponents $\{\beta_p, \gamma_p, \nubar_p\}$ and $\{\tau_a,\sigma_a,\gamma_a, \nubar_a\}$ represent the critical behavior of the SWIR model. 
Because the order parameter does not exhibit critical behavior, $\beta_m=\gamma_m=\nubar_m=0$ and the exponents associated with the diverging behaviors $\gamma_a$ and $\nubar_a$ reduce $\gamma_p$ and $\nubar_p$, respectively. The type of phase transition of the SWIR is mixed-order phase transition (MOT). The critical exponents of the ordinary percolation govern the critical behavior of the MOT  of the SWIR model. 

It may be worth recalling that for the MOTs  observed in other physical models, for instance, the Ising model in one dimension with long-range interaction following the inverse-square law between two spins within the same domain~\cite{mukamel} and  a DNA denaturation model~\cite{dna1,dna2,dna3}, the order parameter does not follow formula (\ref{order}) but jumps discontinuously without exhibiting a critical behavior at a transition point, whereas the susceptibility and the correlation length diverge, as they appear in the second-order transitions. Thus, the MOT in the SWIR model exhibits features similar to those of the above models. 

The paper is organized as follows: In Sec. II, we introduce the SWIR model. In Sec. III, we set up the self-consistency equation to derive the mean-field solution of the order parameter for the epidemic transition on the Erd\H{o}s and R\'enyi (ER) networks. We find that depending on the mean degree of the ER network, different types of phase transition can occur. In Sec. IV, we investigate the properties of those diverse phase transitions. In the final section, a summary and discussion are presented. 

\section{the SWIR  model}

We first define the reactions of the SWIR model as follows: 
\begin{eqnarray}
\rm{S+I}  &\buildrel{\kappa}\over \longrightarrow& {\rm I+I}, \label{si_ii} \\
\rm{S+I} &\buildrel{\mu}\over \longrightarrow& \rm{W+I}, \label{si_wi} \\
\rm{W+I}  &\buildrel{\nu}\over \longrightarrow& \rm{I+I}, \label{wi_ii} \\
\rm{I}  &\buildrel{\lambda}\over \longrightarrow& \rm{R}, \label{i_r} 	
\end{eqnarray}
where $\kappa$, $\mu$, $\nu$ and $\lambda$ denote the contagion rates of the respective reactions between the states of neighboring nodes. For instance, a node in state S can change its state when it contacts with a node in state I to either state I with a probability $\kappa/(\kappa+\mu+\lambda)$, or state W with a probability $\mu/(\kappa+\mu+\lambda)$.  The main use of this SWIR  model is to determine how fast disease spreads on a macroscopic scale with respect to the recovery rate $\lambda$. Thus, without loss of generality, we set $\lambda=1$. On the other hand, when $\mu$ and $\nu$ are much smaller than $\kappa$, the model reduces to the SIR model. Thus, we focus on the opposite limit: the reaction rates of (\ref{si_wi}) and (\ref{wi_ii}) are dominant compared with that of (\ref{si_ii}). Thus, we set $\kappa=0$ and $\nu=1$ for simplicity. The reaction rate $\mu$ serves as a control parameter. For convenience, we  will use the control parameter in an alternative form $r\equiv \mu/(1+\mu)$, which is the reaction probability of (\ref{si_wi}). 

Initially, there exist a single infectious node (seed), the location of which is chosen at random and $N-1$ susceptible nodes. 
We then successively choose which reaction will occur next and when it will occur. The simulation rule is presented in detail in Appendix B. This process is repeated until no infectious nodes remain in the system. This state is called absorbing state. Here we are interested in the behavior of the outbreak size of epidemics, i.e., the fraction of nodes in state R after the system reaches an absorbing state, which serves as the order parameter, denoted as $m$. Moreover, the susceptibility, i.e.,  the fluctuation of the order parameter defined as $\chi_m\equiv N(\langle m^2 \rangle-\langle m \rangle^2)$ averaged over the ensemble is considered as a function of $r$. Using finite-size scaling analysis, we will study phase transitions. \\

\section{Self-consistency equation and physical solutions}

In an absorbing state, each node is in one of three states, the susceptible S,  weakened W and recovered R states. We consider the probability $\ps(\ell)$ that a randomly selected node is in state S after it contacts $\ell$ neighbors in state R.  This probability means that the node remains in state S even though it has been in contact $\ell$ times with those $\ell$ neighbors in state I before they changed their states to R. Thus we obtain
\begin{equation}
\ps(\ell) = (1-r)^{\ell},
\label{p_s}
\end{equation}
where $r$ is the reaction probability given as $r={\mu}/{(1+\mu)}$ with $\kappa=0$ and $\lambda=1$. 
Next, $P_{\rm W}(\ell)$ is similarly defined as the probability that a randomly selected node in state W after it contacts $\ell$ neighbors in state R. The probability $\pw(\ell)$ is given as 
\begin{equation}
\pw(\ell) = \sum_{n=0}^{\ell-1} (1-r)^{n} r (1-w)^{\ell-n},
\label{p_w}
\end{equation}
where $w$ is the probability of the reaction (\ref{wi_ii}),  given as $w=\nu/(\nu+\lambda)={1/2}$. 
Finally, $\pr(\ell)$ is the probability that a node is in state R after it contacts $\ell$ neighbors in state R in the absorbing state. Using the relation $\ps(\ell)+\pw(\ell)+\pr(\ell)=1$, one can determine $\pr(\ell)$ in terms of $\ps$ and $\pw$. 

The order parameter $m(r)$ that a randomly chosen node is in state R after the system falls into an absorbing state is given as
\begin{equation}
m(r)=\sum_{k=1}^{\infty}P_{d}(k) \sum_{\ell=1}^{k} \binom{k}{\ell} q^{\ell}(1-q)^{k-\ell} \pr(\ell),
\label{rho_r}
\end{equation}
where $P_{d}(k)$ is the probability that a node has degree $k$ and $q$ is the probability that an arbitrarily chosen edge leads to a node in state R in the absorbing state. Using the local tree approximation, we define $q_n$ similarly to $q$ but now at the tree level $n$. 
 
The probability $q_{n+1}$ can be derived from $q_n$ as follows:
\begin{equation} \label{eq:sce}
q_{n+1}=\sum_{k=1}^{\infty} \frac{kP_d(k)}{\langle k\rangle} \sum_{l=0}^{k-1} \binom{k-1}{\ell} q_n^{\ell}(1-q_n)^{k-1-\ell} \pr(\ell)\equiv f(q_n),
\end{equation}
where the factor $kP_d(k)/{\langle k\rangle}$ is the probability that a node connected to a randomly chosen edge has degree $k$. As a particular case, when the network is an ER network having a degree distribution that follows the Poisson distribution, i.e., $P_d(k)=\langle k \rangle^k e^{-\langle k \rangle}/k!$, where $\langle k \rangle =\sum_k kP_d(k)$ is the mean degree, the function $f(q_n)$ is reduced as follows:
\begin{widetext}
\begin{equation} \label{eq:f_q}
f(q_n)=1-e^{-rq_n \langle k \rangle}+\dfrac{r}{1-2r}e^{-q_n \langle k \rangle/2}-\dfrac{r}{1-2r}e^{-rq_n \langle k \rangle}.
\end{equation}
\end{widetext}
 
Eq.~(\ref{eq:sce}) reduces to a self-consistency equation for $q$ for a given reaction rate $r$ in the limit $n\to \infty$. Once we obtain the solution of $q$, we can obtain the outbreak size $m(r)$ using Eq.~(\ref{rho_r}). For ER networks, however, $m(r)$ becomes equivalent to $q$ so that the solution of the \self equation Eq.~(\ref{eq:sce}) yields the order parameter.  We remark that the method we used is similar conceptually to those used in previous studies of epidemic spreading on complex networks~\cite{hasegawa, dodds,bizhani, chung,janssen_spinodal}.

For convenience, we define a function $G(m)\equiv f(m)-m$. Using formula (\ref{eq:f_q}), we approximate $G(m)$ in the limit $m\to 0$ as 
\begin{equation} \label{eq:G_q}
G(m)=a m+b m^2+ c m^3+O(m^4), 
\end{equation}
where 
\begin{eqnarray}
a&=&\frac{1}{2}(r-r_a)\langle k \rangle, \label{a_c}\\
b&=&\frac{1}{4}r(r_b-r)\langle k \rangle^2, \label{b_c}\\
c&=&\frac{1}{12}r(r-r_c^+)(r-r_c^-)\langle k \rangle ^3 \label{c_c} 
\end{eqnarray}
with $r_a=2/\langle k \rangle$, $r_b=1/2$ and $r_c^{\pm}=(1\pm \sqrt{5})/2$. Because $r_c^-<0$, $c$ can change sign only across $r_c^+$ in the range $0< r_c^+<1$. However, because $G(m) \to -\infty$ as $q\to \infty$, we limit our investigation to the range $r< r_c^+$ hereafter, so that $c$ is always negative. For convenience, we neglect the higher order terms and redefine $G(m)$ as
\begin{equation}
G(m)=am+bm^2+cm^3.
\label{eq:h_q}
\end{equation} 

Depending on the relative magnitude between $a$ and $b$, various solutions of the self-consistency equation $G(m)=0$ can exist. However, we need to check whether those solutions are indeed physically relevant in the steady state when we start epidemic dynamics from the given initial condition. We set up the stability criterion as follows: We impose a small perturbation to the steady state solution $q^*$ of Eq.~(\ref{eq:sce}). Then we can obtain the recursive equation as 
\begin{equation}
q^{*}+\delta q_{n+1}\approx f(q^{*})+\frac{df}{dq}\Big|_{q=q^*} \delta q_n,
\end{equation}
which leads to 
\begin{equation}
\eta \equiv \dfrac{\delta q_{n+1}}{\delta q_{n}}=\frac{df}{dq}\Big|_{q=q^*}.
\end{equation}
If $\eta < 1$ ($> 1$), then the steady state solution $q^*$ is stable (unstable).  

\section{Phase transitions} 
The equation of state in the steady state can be obtained using $G(m)=0$. From Eq.~(\ref{eq:h_q}), there exist one trivial solution $m=0$ and two non-trivial solutions $m=m_d$ and $m_u$, where  
\begin{eqnarray} 
m_d(r)=-\dfrac{b}{2c} - \sqrt{ \dfrac{b^{2}}{4c^{2}}-\dfrac{a}{c}}, \label{eq:nontrivialSol1} \\
m_u(r)=-\dfrac{b}{2c} + \sqrt{ \dfrac{b^{2}}{4c^{2}}-\dfrac{a}{c}}. \label{eq:nontrivialSol2}
\end{eqnarray}
Particularly, when $b^2-4ac=0$, $m_d=m_u$, which is denoted as $m_*$. 
Depending on the relative magnitude between $r_a=2/\langle k \rangle$ and $r_b=1/2$, which determines the signs of $a$ and $b$,  diverse types of non-trivial solutions of $G(m)=0$ exist. Thus, we consider the cases $\langle k \rangle > 4$, $\langle k \rangle=4$ and $\langle k \rangle < 4$, separately. 

\begin{figure}[h]
\centering
\includegraphics[width=0.98\linewidth]{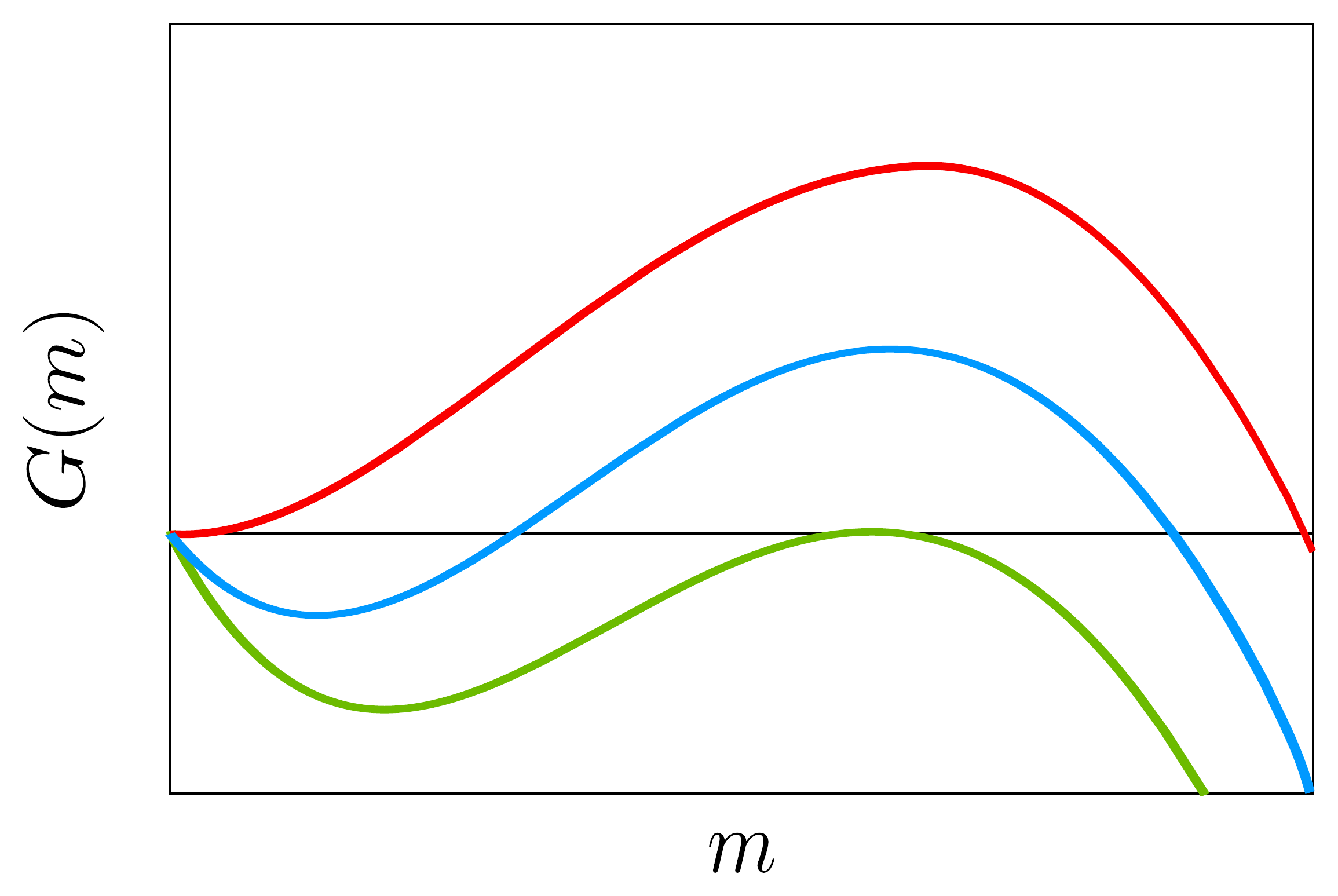}
\caption{For $\langle k \rangle > 4$, schematic plot of $G(m)$ versus $m$ for fixed reaction rates $r=r_*$ (bottom, green), $r_*< r < r_a$ (middle, blue), and $r=r_a$ (top, red). $G(m)$ becomes zero at $m=0$, $m_d$ and $m_u$, which are determined using Eqs.~(\ref{eq:nontrivialSol1}) and (\ref{eq:nontrivialSol2})}.
\label{fig1}
\end{figure}

\begin{figure}[h]
\centering
\includegraphics[width=0.98\linewidth]{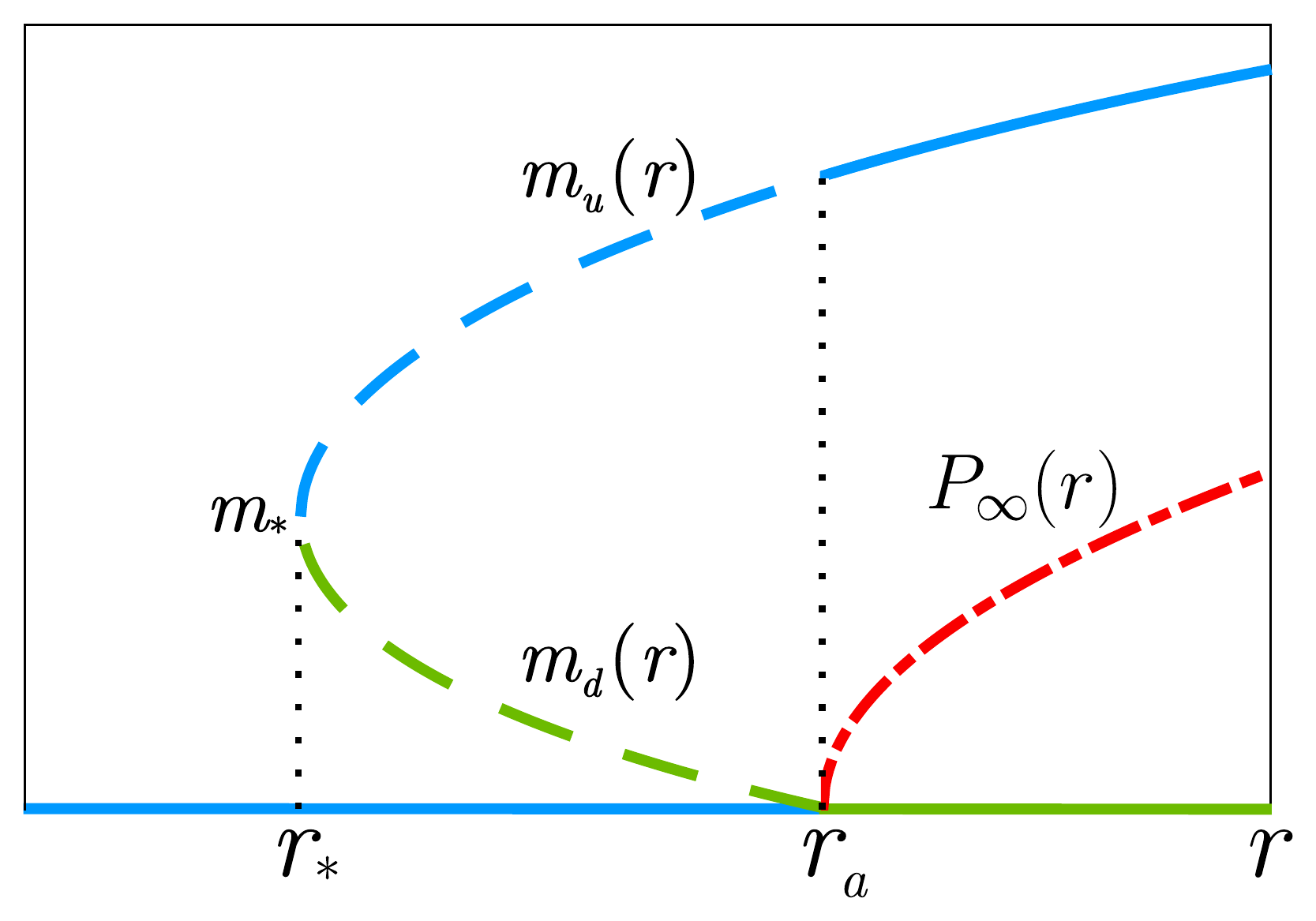}
\caption{Schematic plot of $m(r)$ for $\langle k \rangle > 4$. Stable solutions of $m(r)$ are represented by blue (solid or dashed) curve and line,  whereas unstable solutions are done by green (solid or dashed) curve and line. Physically accessible states are indicated by solid lines, whereas inaccessible states are indicated by dashed lines. The probability $\pinf(r)$ is indicated by dashed-dotted curve. The order parameter $m(r)$ jumps from $m=0$ to $m_u(r)$ at $r_a$ with the probability $\pinf(r)$.}
\label{fig2}
\end{figure}

\subsection{For $\langle k \rangle > 4$}
\begin{figure}[h]
\centering
\includegraphics[width=0.98\linewidth]{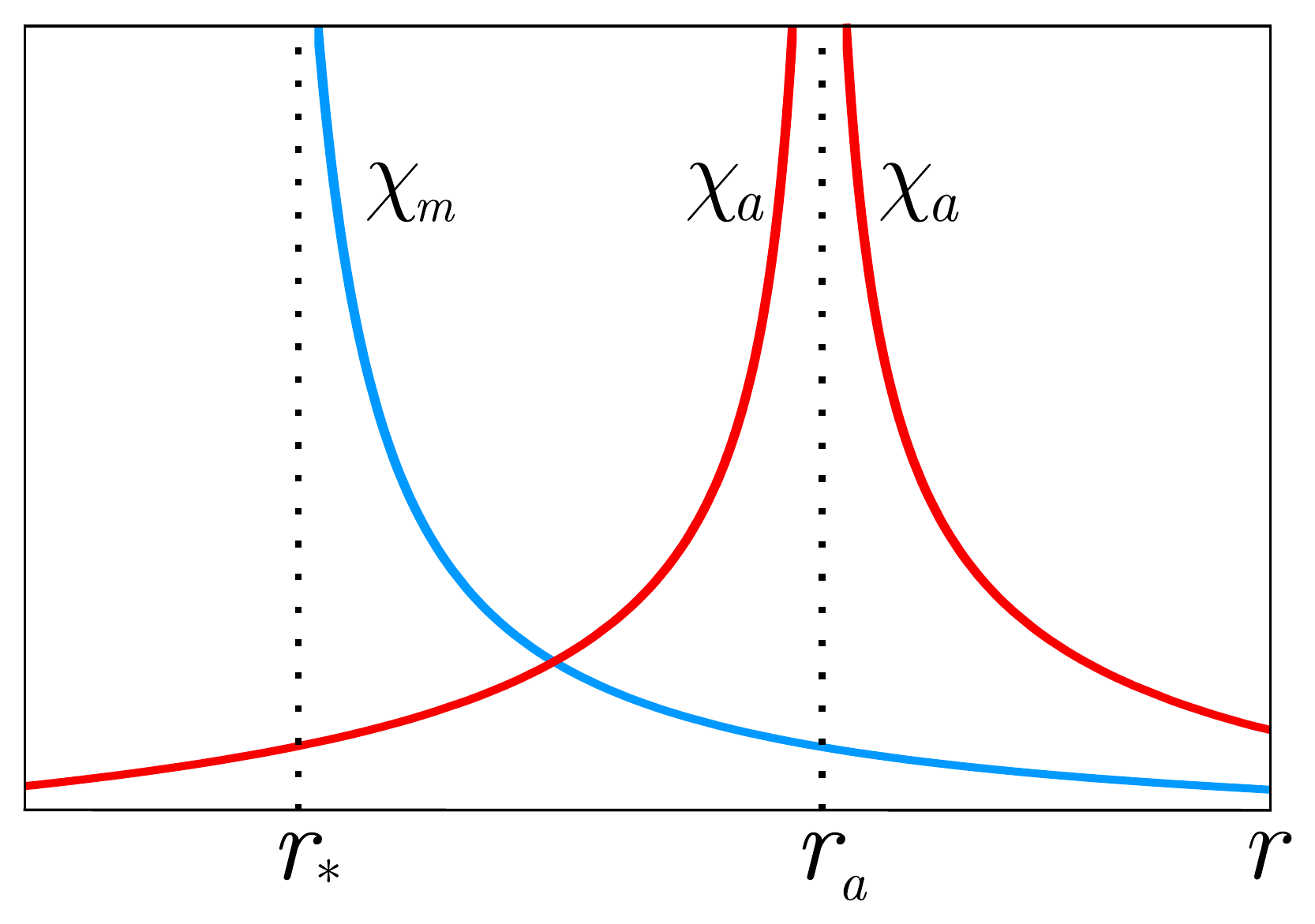}
\caption{Schematic plot of the susceptibilities $\chi_m$ and $\chi_a$ defined in the text as a function of $r$ in the thermodynamic limit. They show peaks at $r_*$ and $r_a$, respectively. We remark that $\chi_m$ does not diverge, but $\chi_a$ diverges at the transition point $r_a$}
\label{fig3}
\end{figure}

When $\langle k \rangle > 4$, $r_a < r_b$. The behavior of $G(m)$ as a function of $m$ is schematically shown in Fig.~\ref{fig1} and the solution $m$ of $G(m)=0$ as a function of $r$ is schematically shown in Fig.~\ref{fig2}. There are several mathematical solutions: the physically relevant solution of the order parameter is indicated by solid line for $r < r_a$ and by solid curve for $r > r_a$. At $r_a$, the order parameter jumps to the extent of $m(r_a)$. The details are described as follows: 

i) For  $r < r_* < r_a$, there exists one stable solution $m=0$. Recall that $r_*$ is the solution of the equation $b^2-4ac=0$. 

ii) At $r=r_* < r_a$, there exist one trivial solution $m=0$ and one nontrivial solution $m=m_* > 0$, where $m_*=-b/(2c)$.  The solution $m_*$ is not accessible in the thermodynamic limit because there exists one stable solution $m=0$. The probability $\pinf(r)$ that an infinite outbreak occurs in a given sample is zero in the thermodynamic limit. However, in finite systems, the probability $\pinfn(r)$ that an outbreak of size $O(N)$ occurs can be nonzero even for $r < r_a$ (see also Fig.~\ref{fig8}). Thus, the solution $m=m_*$ could be observed in finite systems.  We remark that the susceptibility of the order parameter $\chi_m=N(\langle m^2 \rangle-\langle m \rangle^2)$ diverges at $r_*$ as shown schematically in Fig.~\ref{fig3}.

\begin{figure}[t]
\centering
\includegraphics[width=0.99\linewidth]{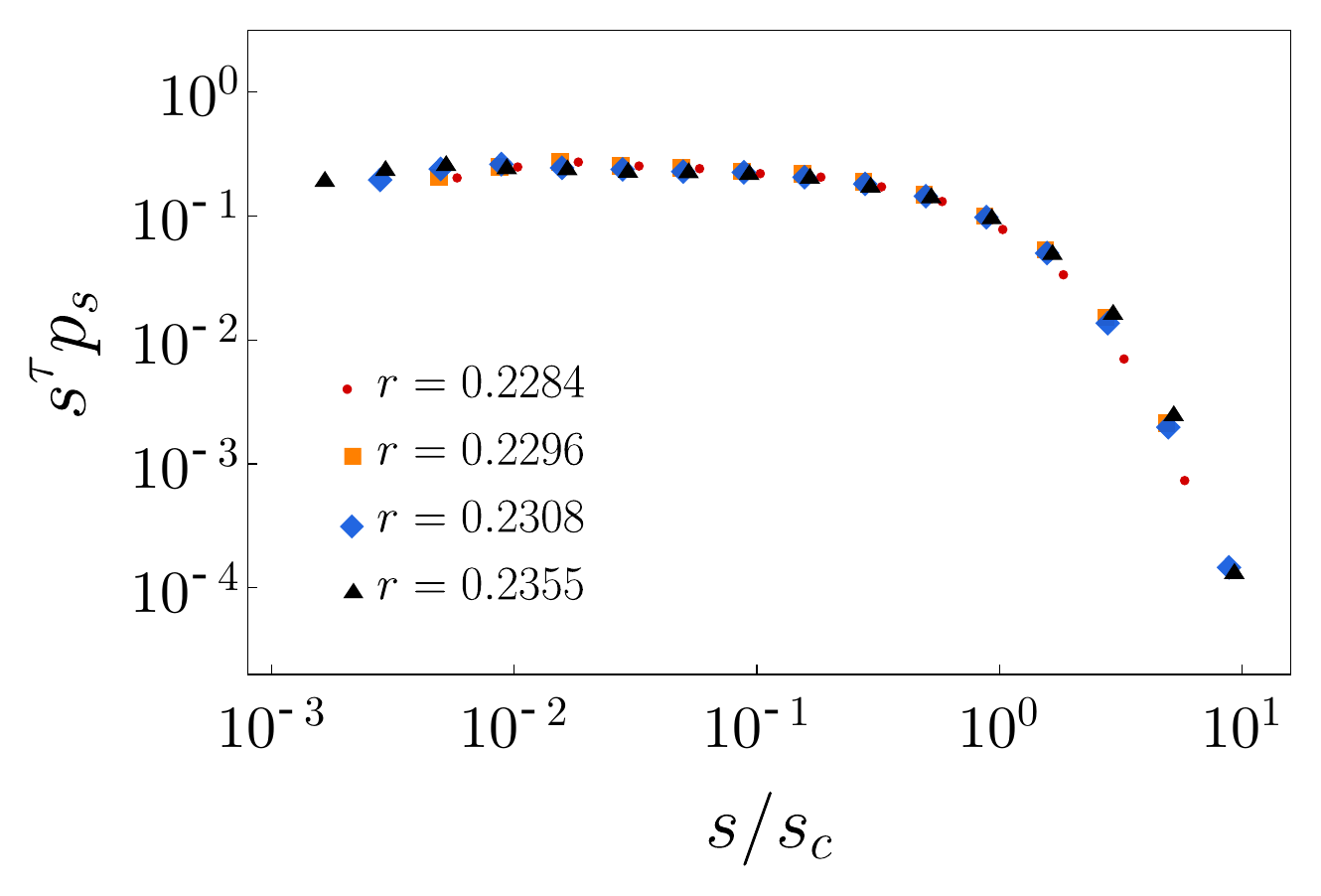}
\caption{Scaling plot of the outbreak size distribution $s^{\tau_a}p_s(r)$ versus $s/s_c$ for several values of $r < r_a$, in which $ \tau_a=1.5 $ and $s_c\sim (r_a-r)^{-1/\sigma_a}$ with $\sigma_a=0.5$ are used.}
\label{fig4}
\end{figure}

\begin{figure}[t]
\centering
\includegraphics[width=0.99\linewidth]{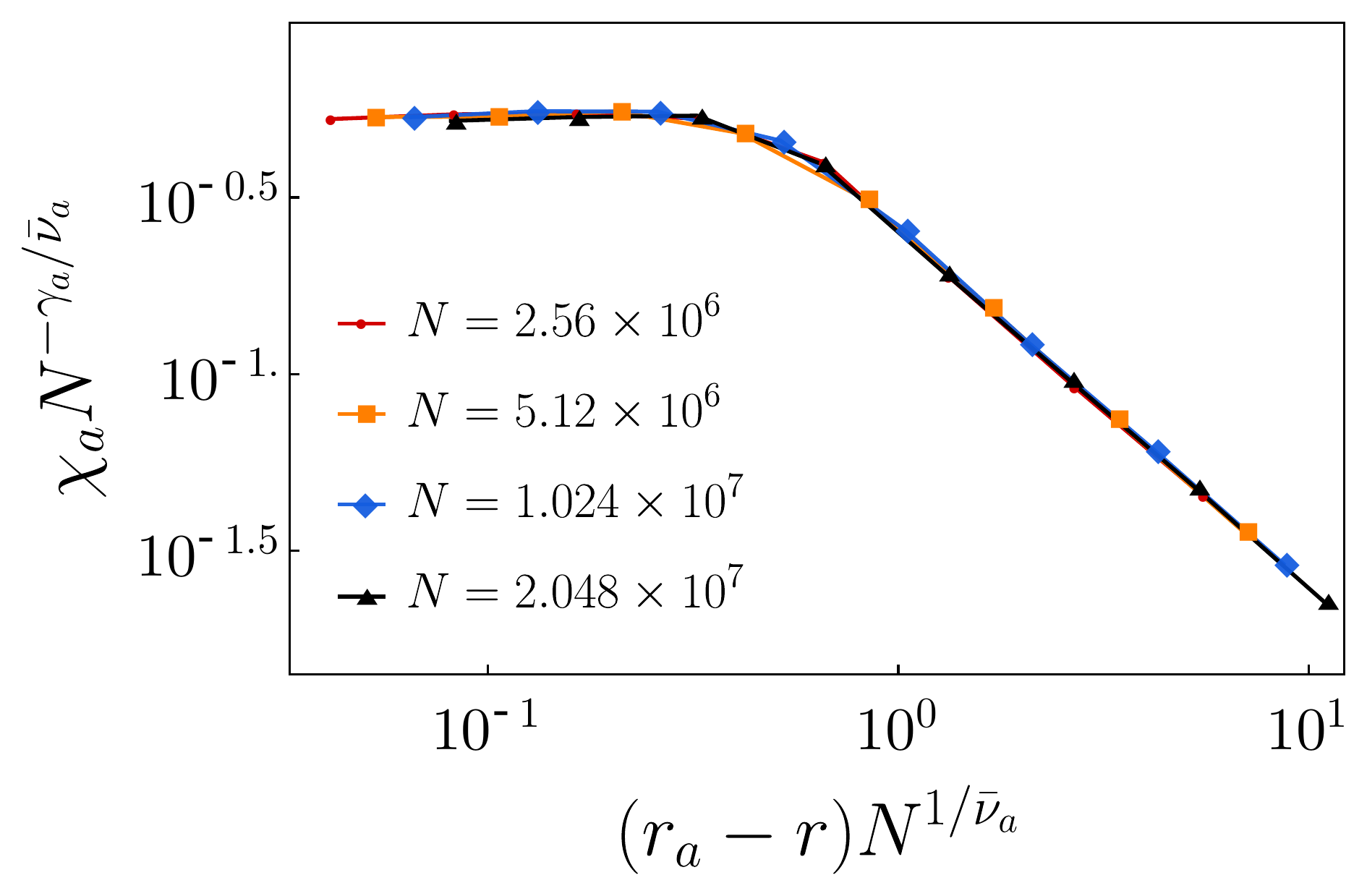}
\caption{Scaling plot of the susceptibility $\chi_a$ versus the reaction rate $r < r_a$ in the form $\chi_a N^{-\gamma_a/\nubar_a}$ and $(r_a-r)N^{1/\nubar_a}$, respectively. Data are obtained from systems of different system sizes $N$. With the choice of $\gamma_a=1$ and $\nubar_a=3$, data from the different system sizes are well collapsed  onto a single curve.}
\label{fig5}
\end{figure}

iii) When  $r_* < r  < r_a$, there exist one trivial and stable solution $m=0$ and two nontrivial solutions $m_d(r)$ and $m_u(r)$. The solution $m_d$ is unstable but $m_u$ is stable. Because the initial density of infectious seeds $\ro=0$ and $\pinf(r)=0$ in this interval, the solution $m_u$ is inaccessible and unphysical. However, in finite systems, the order parameter can have the solution $m_u(r)$ with the probability $\pinfn(r)$ (see also Fig.~\ref{fig8}).

iv)  At $r=r_a$, there exist one trivial solution $m=0$ and one nontrivial solution $m=m_u$ as the case iii). Finite and infinite outbreaks can occur. The size distribution of finite outbreaks around $r_a$ follows a power law with an exponential cutoff as $p_s(r)\sim s^{-\tau_a}{\rm exp}(-s/s_c)$, where $\tau_a \approx 1.5$ and $s_c\sim |r-r_a|^{-1/\sigma_a}$ with $\sigma_a\approx 0.5$ (Fig.~\ref{fig4}). The mean size $\langle s \rangle$ of finite outbreaks exhibits a diverging behavior, which is another susceptibility defined as $\chi_a \equiv \langle s \rangle =\sum sp_s$, as $\sim (r_a-r)^{-\gamma_a}$. From the scaling relation, it follows that $\gamma_a=(2-\tau_a)/\sigma_a \approx 1$. In finite systems, the susceptibility diverges as $\chi_a\sim N^{\gamma_a/\nubar_a}g(|r-r_a|N^{1/\nubar_a})$ (Fig.~\ref{fig5}), where $\nubar_a$ is the exponent associated with the correlation size of finite outbreaks.  We confirm that the measured value $\gamma_a$ satisfies the scaling relation $\gamma_a=(2-\tau_a)/\sigma_a$. The exponent $\nubar_a\approx 3$ is obtained. However, $\chi_m(r_a)$ does not diverge. The probability $\pinf(r_a)=0$ but $\pinfn(r_a)\ne 0$ in finite systems. Thus there can exist infinite outbreaks of size $Nm_u(r)$ with the probability $\pinfn(r_a)$ in finite systems. 

v) For $r > r_a$, there exist one unstable solution $m=0$ and one stable nontrivial solution $m=m_u$. Thus, the system can be in pandemic state to the extent of $m=m_u$ with the probability $\pinf(r)$. With the remaining probability $1-\pinf(r)$, the system remains in the state $m=0$. 
The probability $\pinf(r)$ is equivalent to the spanning probability of percolation~\cite{grassberger_nphy,branching}, which is given as $\sim (r-r_a)^{\beta_p}$, where $\beta_p$ is the exponent for the order parameter of the ordinary percolation transition, which is known as $\beta_p=1$ for ER networks. When an infinite outbreak occurs, the order parameter $m(r)$ behaves as  $m(r)-m_u(r_a)\sim (r-r_a)$. However, the susceptibilities $\chi_m$ both at both $(r_a, 0)$ and $(r_a, m_u(r_a))$ do not diverge. Critical behavior of $\chi_a$ occurs at $(r, m)=(r_a, 0)$ owing to the singular behavior of $\pinf(r)$. 

\begin{figure}[t]
\centering
\includegraphics[width=0.99\linewidth]{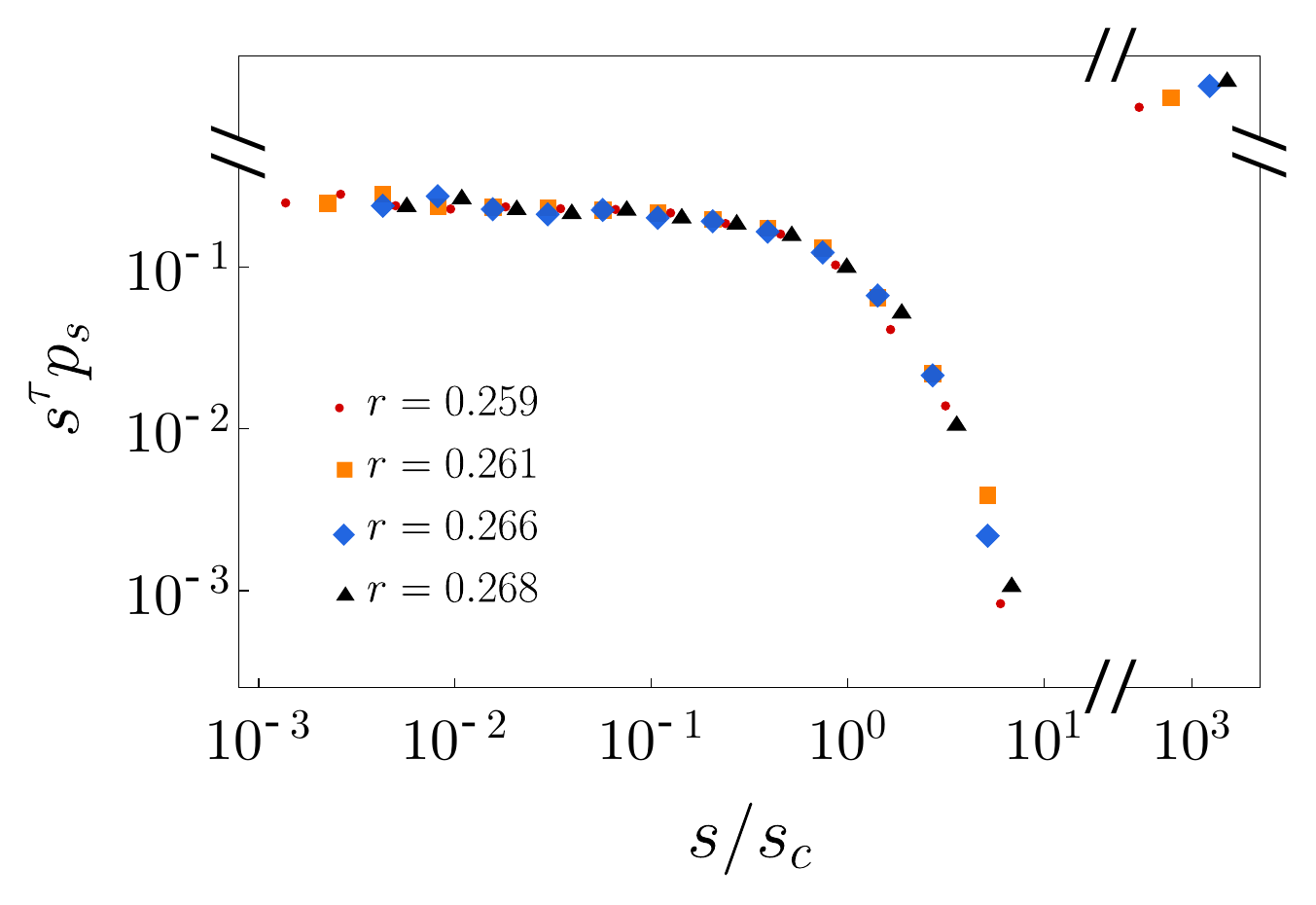}
\caption{Scaling plots of the outbreak size distribution versus $s$ for several values of $r > r_a$, in which $ \tau_a=1.5 $ and $s_c\sim (r-r_a)^{-1/\sigma_a}$ with $\sigma_a= 0.5$ are used. Data of macroscopic-scale outbreak sizes appear away from the curves of finite outbreaks. }
\label{fig6}
\end{figure}

\begin{figure}[t]
\centering
\includegraphics[width=0.99\linewidth]{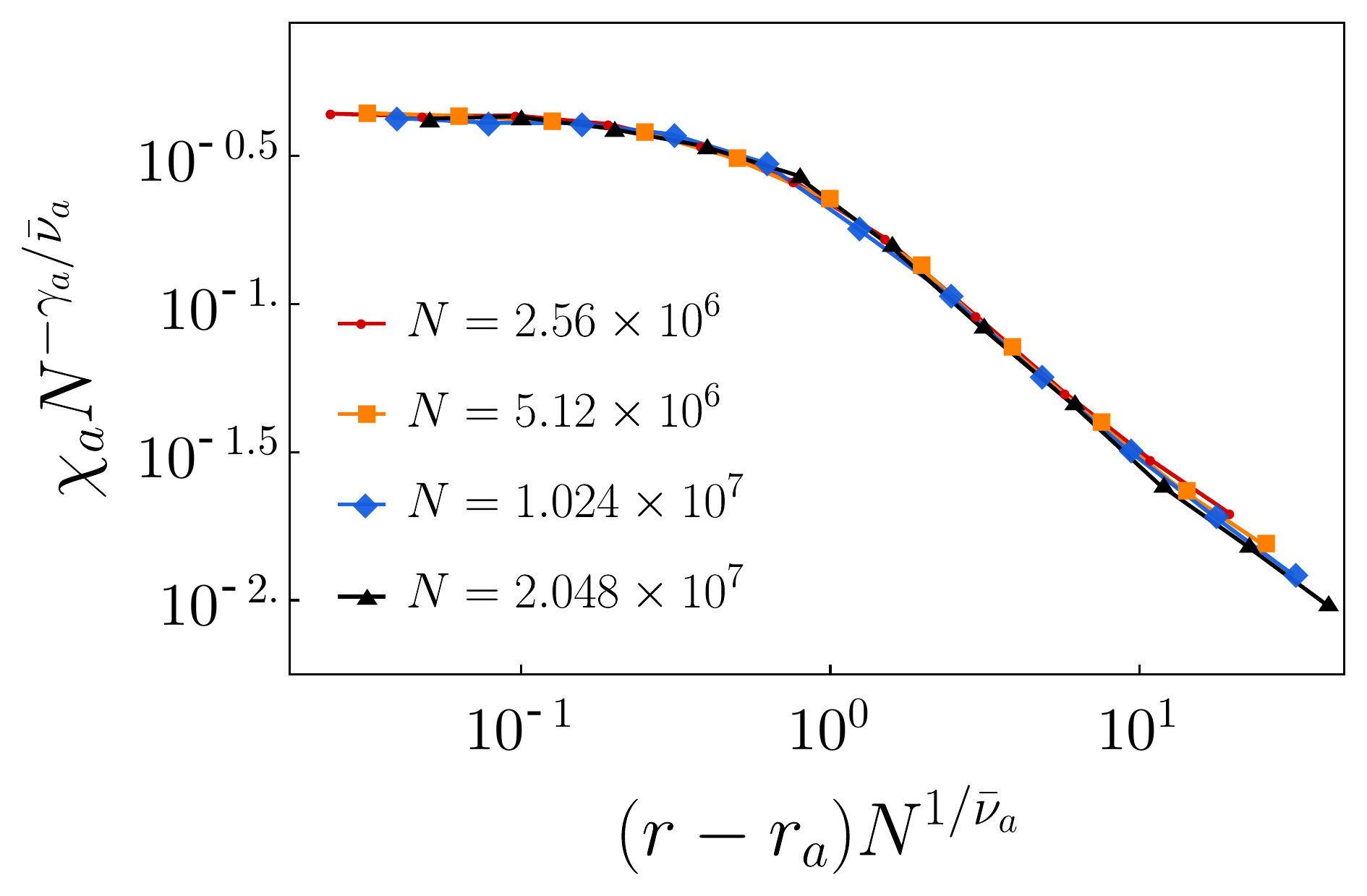}
\caption{Scaling plot of the susceptibility $\chi_a$ versus the reaction rate $r$ in the range $r > r_a$ for different system sizes $N$. With the choice of $\gamma_a=1$ and $\nubar_a=3$, data from different system sizes are well collapsed on a single curve.}
\label{fig7}
\end{figure}

\begin{figure}[t]
\centering
\includegraphics[width=0.98\linewidth]{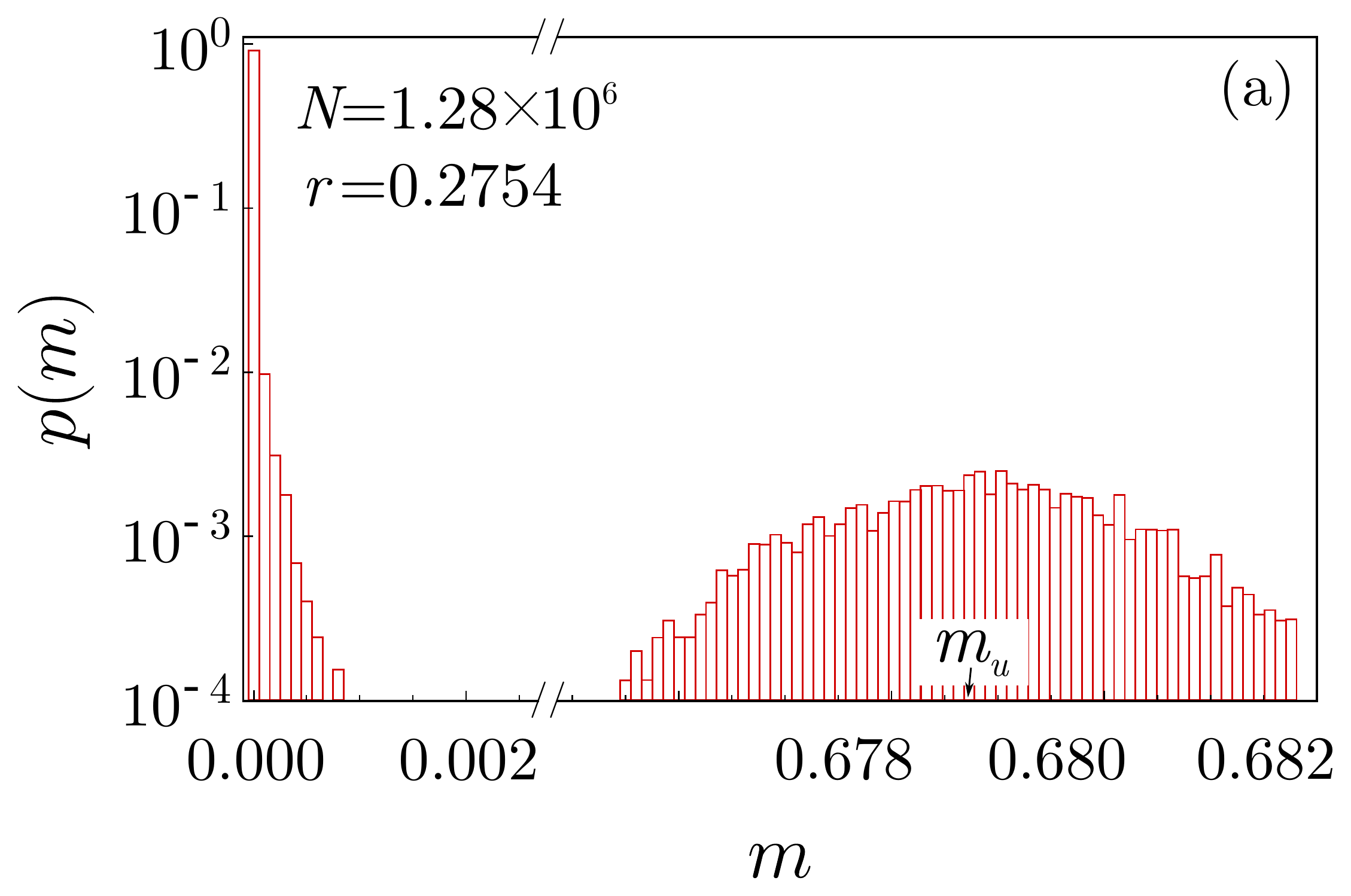}
\includegraphics[width=0.98\linewidth]{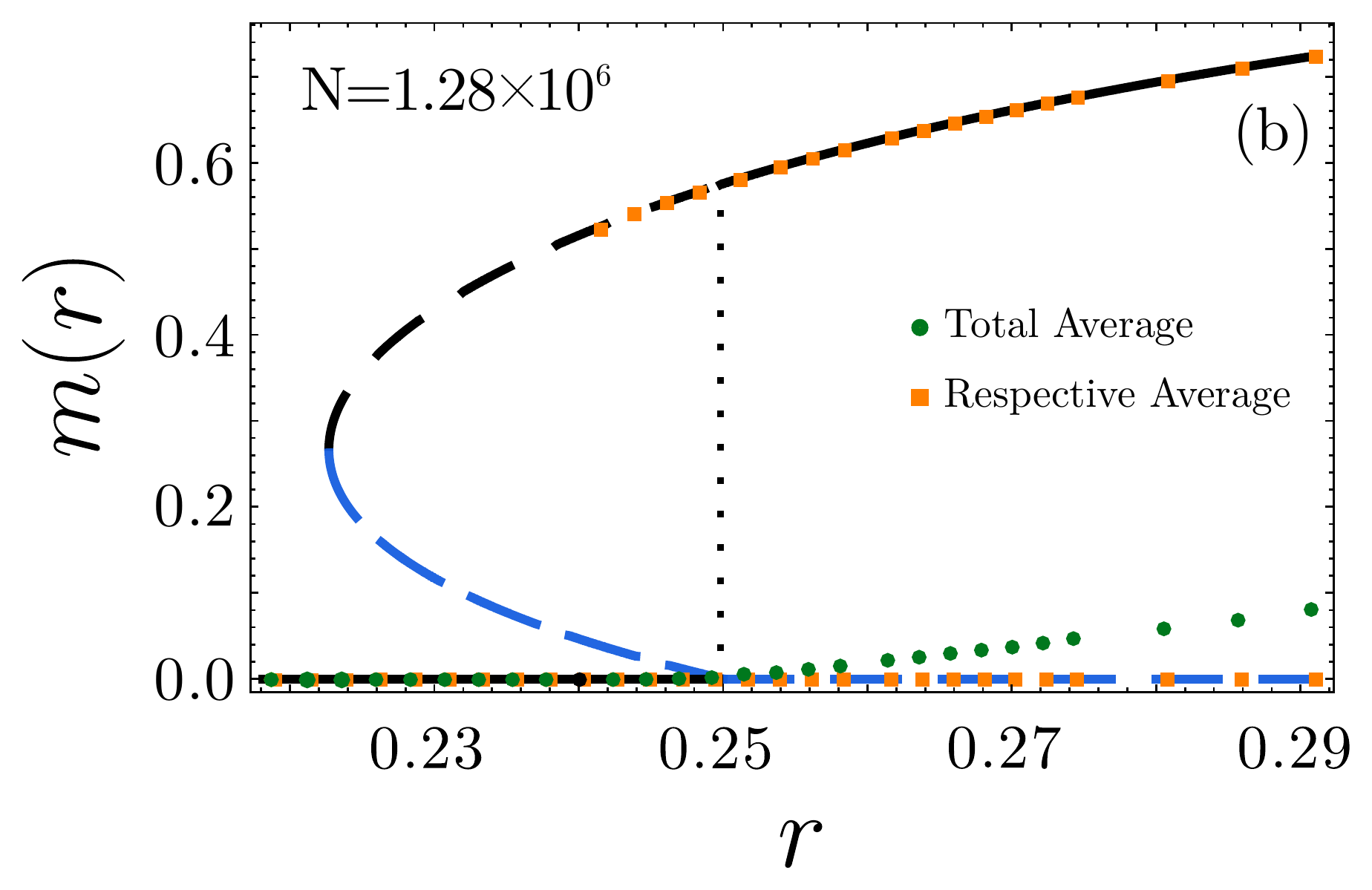}
\caption{(a) Plot of the fraction $p(m)$ of the samples having $m$ versus $m$. The distribution is separated into the two curves composed of finite and infinite outbreak samples. (b) Plot of numerical data of $m(r)$ versus $r$ on the theoretical curve shown in Fig.~2.  Data are obtained in two different ways, averaged over all samples (green $\bullet$), over respective finite-outbreak and infinite-outbreak samples (orange $\square$).}
\label{fig8}
\end{figure}

In finite systems, the distribution of finite outbreak sizes for $r > r_a$ is similar to that for $r < r_a$ as $p_s(r)\sim s^{-\tau_a}{\rm exp}(-s/s_c)$, where $\tau_a \approx 1.5$ and $s_c\sim (r-r_a)^{-1/\sigma_a}$ with $\sigma_a\approx 0.5$ (Fig.~\ref{fig6}). The mean size $\langle s \rangle$ of finite outbreaks exhibits a scaling behavior, which is the susceptibility $\chi_a \equiv \langle s \rangle =\sum sp_s$, as $\sim N^{\gamma_a/\nubar_a}$ (Fig.~\ref{fig7}), where $\nubar_a$ is the exponent associated with the correlation size.  It turns out to be that $\nubar_a=\nubar_p$.  

Here we discuss finite-size scaling behavior. We choose $\langle k \rangle=8$ for simulations, thus the transition point is located at $r_a=1/4$. 
In Fig.~\ref{fig8}(a), we examine  the probability $p(m)$ that at a certain $r=0.2754 > r_c$ the system has outbreak size $m$. We find that there exist two peaks: one peak at $m=0$ and the other at $m_u(r) > 0$. This behavior occurs for any $r$-value above $r_c$, even though their peak heights change depending on $r$.  This result supports the idea that outbreaks need to be categorized into two types: finite and infinite outbreaks. The order parameter is obtained by taking two different types of ensemble average: i) over all samples and ii) over respective samples of finite and infinite outbreaks. The numerical values of $m(r)$ obtained from the two types of averages are denoted as $m_t(r)$ and $m_u(r)$, respectively. As shown in Fig.~\ref{fig8}(b), $m_t(r)$ (green $\bullet$) increase continuously with $r$. However, data of $m_u(t)$ (orange $\square$) locate on the theoretical curve $m_u(r)$, respectively. The data lying on the line $m\approx 0$ is the average value over finite outbreaks, which is almost zero. 

\begin{figure}[ht]
\centering
\includegraphics[width=0.99\linewidth]{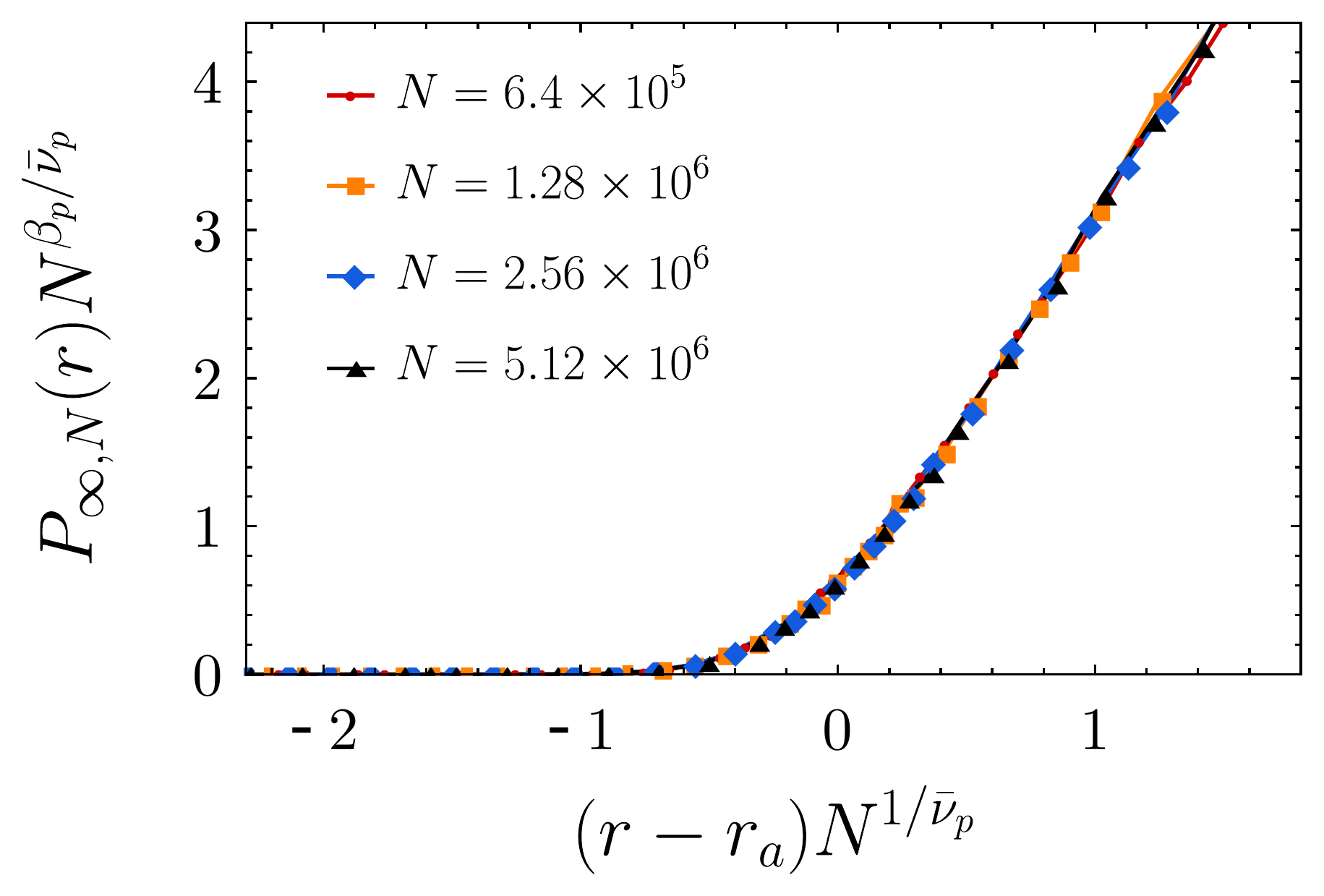}
\caption{Scaling plot of rescaled outbreak probability $\pinfn(r)N^{\beta_p/\nubar_p}$ that an infinite outbreak occurs in a certain sample versus rescaled reaction rate $\Delta r N^{1/\nubar_p}$. With the choice of known values $ \beta_p = 1 $ and $\nubar_p=3$, the data are well collapsed onto a single curve.}
\label{fig9}
\end{figure}

Next, we examine numerically the probability $\pinf(r)$ that an infinite outbreak occurs in a certain sample, which is proposed as $\pinf(r) \sim (r-r_a)^{\beta_p}$ with $\beta_p=1$. By applying finite-size scaling analysis, the probability $\pinfn(r)$ in finite systems can be written in the scaling form of $\pinfn \sim N^{-\beta_p/\nubar_p}g((r-r_a)N^{1/\nubar_p})$, where $g(x)$ is a scaling function. Indeed in Fig.~\ref{fig9}, we find that data for systems of different system sizes $N$ are well collapsed onto a single curve. From this figure, we find that infinite outbreaks rarely occur for $r \ll r_a$, and the probability gradually increases as $r$ approaches $r_a$ in finite systems. As argued in Ref.~\cite{grassberger_nphy}, $\pinf(r)$ is actually the order parameter of the SIR transition, and is the probability to create a critical branching tree of size $O(N^{2/3})$ ~\cite{branching}. 

\begin{figure}[h]
\centering
\includegraphics[width=0.99\linewidth]{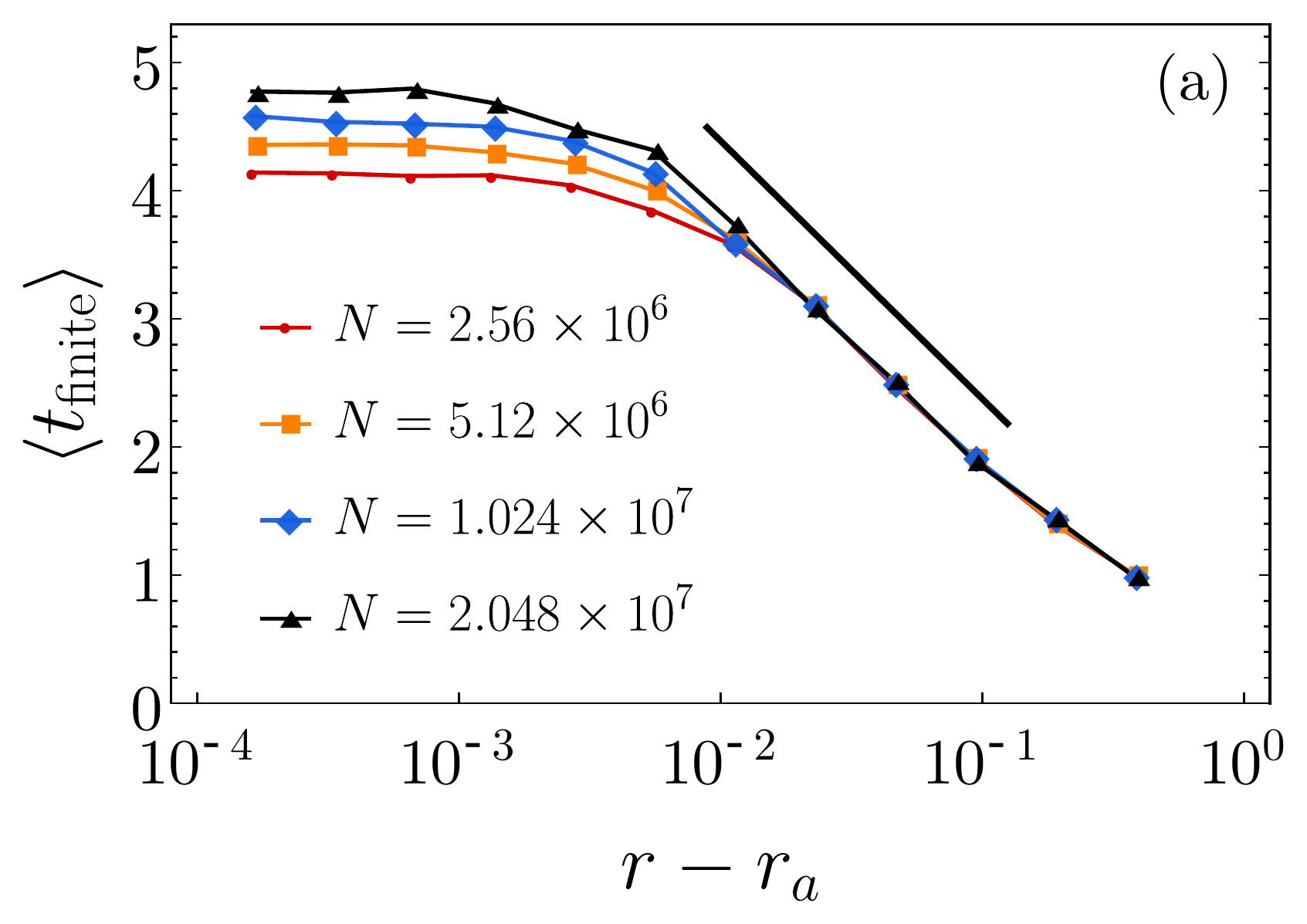}
\includegraphics[width=0.99\linewidth]{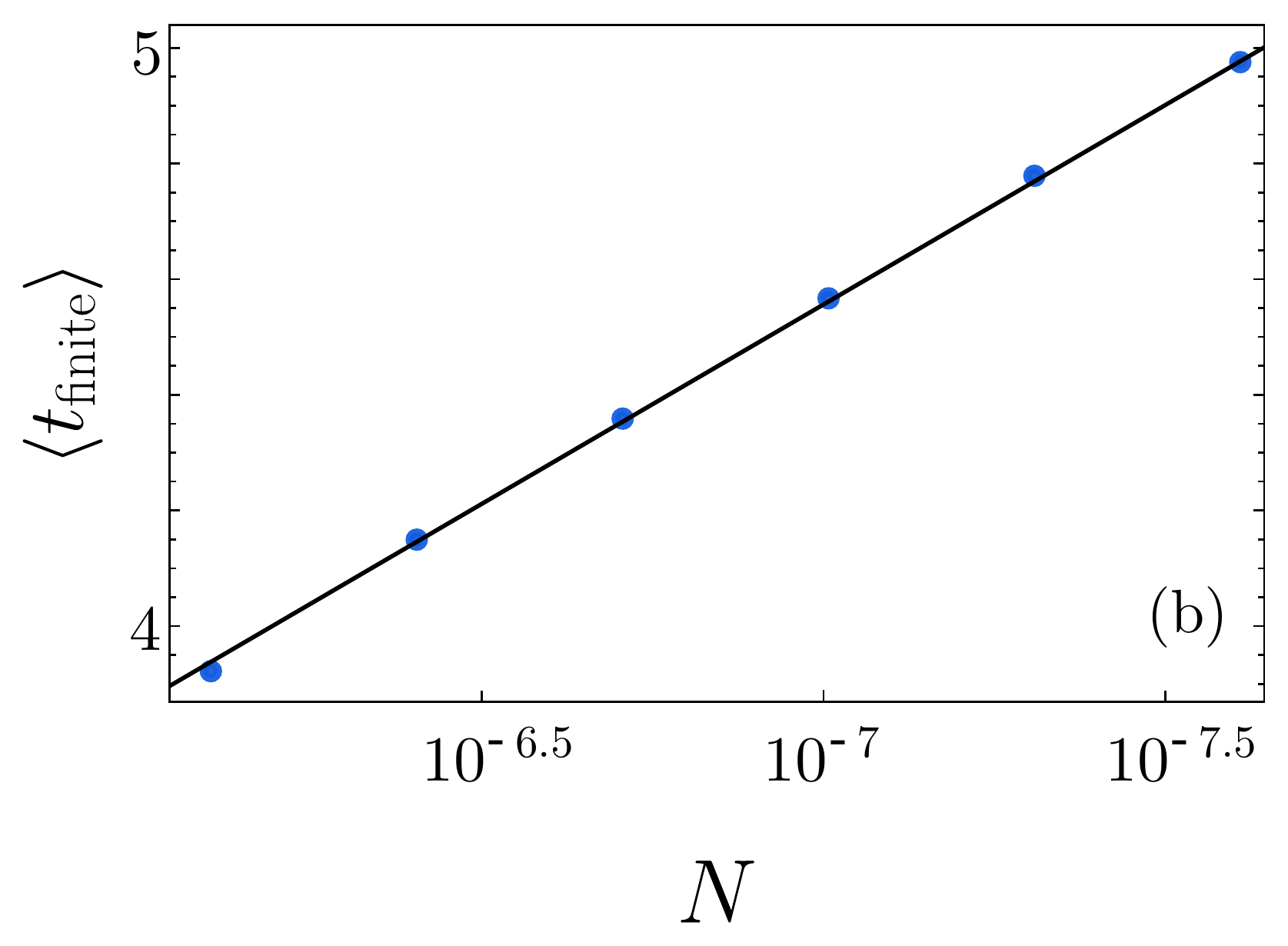}
\caption{Plot of the mean outbreak time of finite outbreaks $\langle t_{\rm finite}\rangle$ as a function of (a) $\Delta r=r-r_a$ and (b) $N$ on semilogarithmic scales.}
\label{fig10}
\end{figure}

We investigate the mean outbreak time of finite outbreaks of size $s$. The outbreak time is the continuous time required to reach an absorbing state. We explain how to calculate a continuous outbreak time in Appendix B. Numerically it is found that $t_{\rm finite} \sim s^{0.5}$. Using the outbreak size distribution $p_s(r)$ and the relations $p_sds = p_tdt$ and $s \sim t^2$, we obtain that $p_t(r)\sim t^{-2\tau_a+1}f(t^2/(r-r_a)^{-1/\sigma_a})$. Thus, the mean outbreak time for finite outbreaks scales as $\langle t_{\rm finite}\rangle \sim -\ln (r-r_a)$ for $r > r_a$ (Fig.~\ref{fig10}(a)) and as $\langle t_{\rm finite} \rangle \sim \ln N$ at $r=r_a$  (Fig.~\ref{fig10}(b)). 

We remark that the distribution $p_tdt$ can be interpreted as the probability that a spreading epidemic terminates between $t$ and $t+dt$. Then, the surviving probability of the epidemic dynamics surviving up to the time step $t$ is obtained as $q_t=\int p_{t^{\prime}} dt^{\prime}$, which is denoted as $q_t \sim t^{-\delta_d}$ following the convention used in the theory of the absorbing phase transition and thus $\delta_d=2\tau_a-2=1$. Next, the number of nodes (denoted as $u(t)$) that change their state to R at step $t$ averaged over the surviving configurations is obtained by $ds(t)/dt$, which is conventionally denoted as $u(t)\sim t^{\eta_d+\delta_d}$. Thus, $\eta_d=0$.  The exponent values $\eta_d=0$ and $\delta_d=1$ are equivalent to the mean field values of the directed percolation universality class~\cite{absorbing}. 

\begin{figure}[h]
\centering
\includegraphics[width=0.99\linewidth]{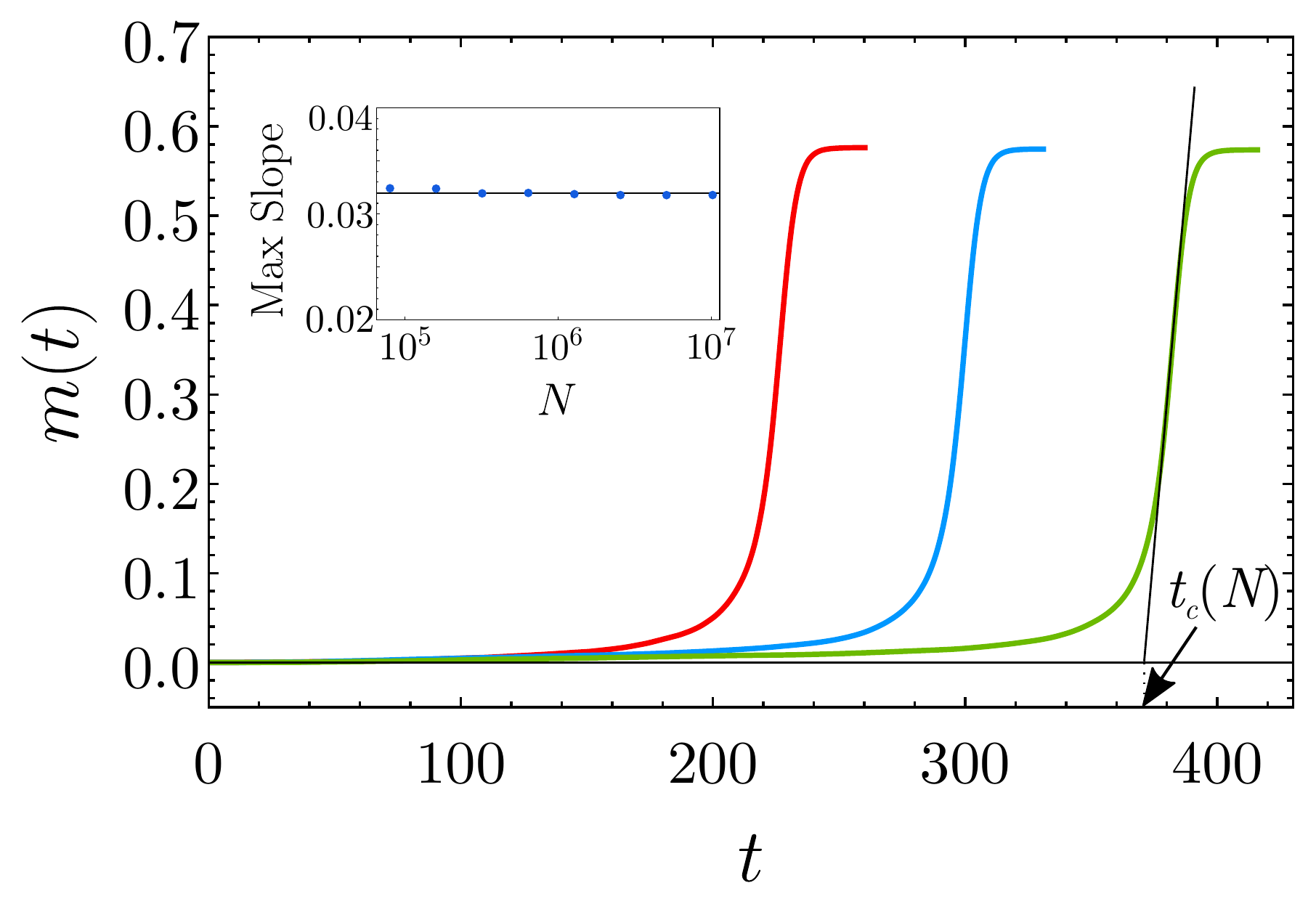}
\caption{Plot of temporal evolution of the order parameter as a function of time step $t$ for system size $N/10^6=2^8, 2^9$ and $2^{10}$ from left to right for infinite outbreaks. Inset: Plot of the mean maximum slope versus $N$. The slopes show independent behavior of $N$, indicating that the increase rate of the infinite outbreak size is independent of the system size.}
\label{fig11}
\end{figure}

\begin{figure}[h]
\centering
\includegraphics[width=0.99\linewidth]{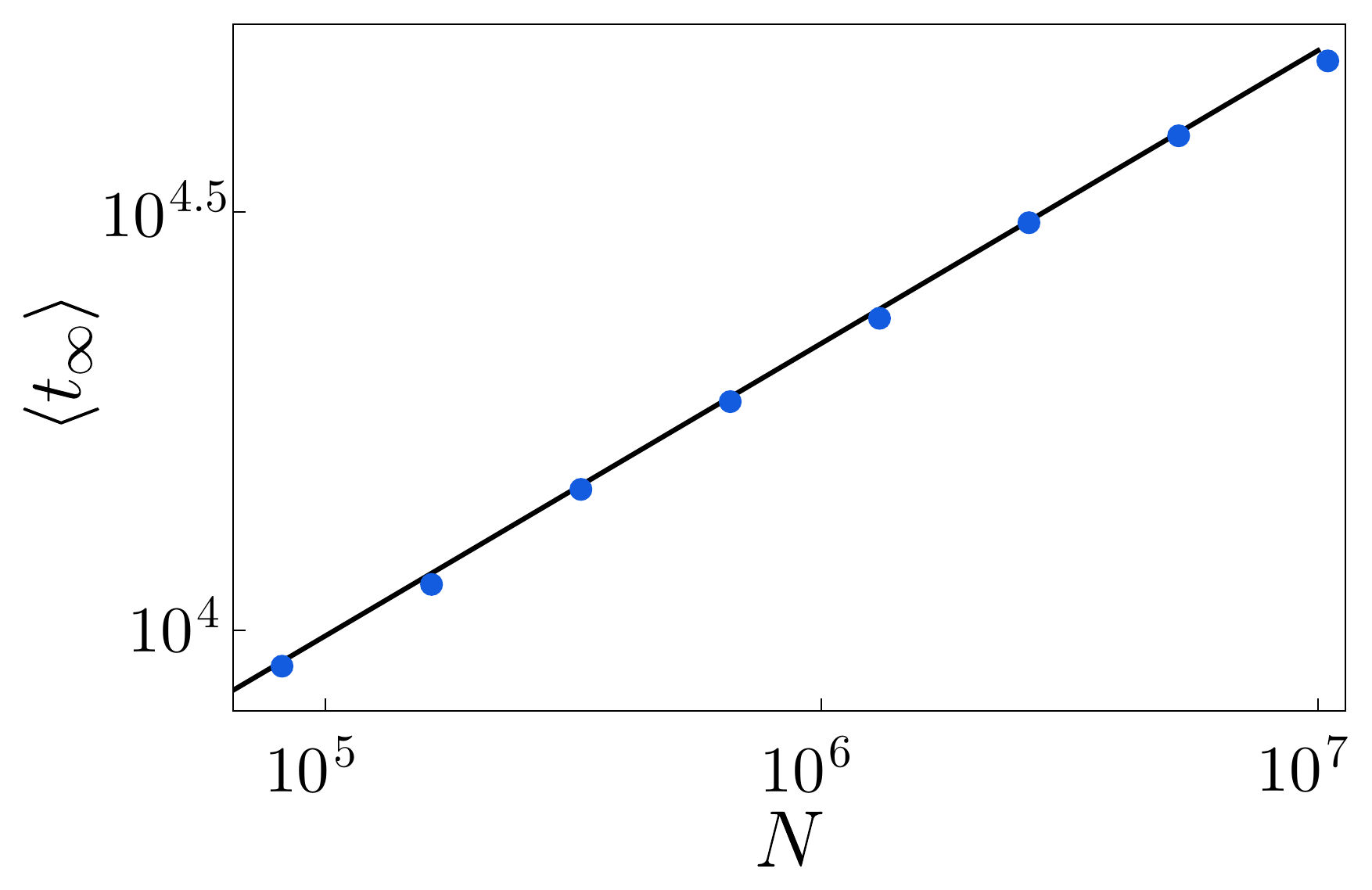}
\caption{Plot of the mean outbreak time of infinite outbreaks $\langle t_{\infty} \rangle$ versus $N$ on double logarithmic scale. The guideline has a slope of $0.35$.}
\label{fig12}
\end{figure}

The mean outbreak time of infinite outbreaks differs from that of finite outbreaks. To study the mean outbreak time of infinite outbreaks, we plot the temporal evolution of the order parameter as a function of time for several system sizes in Fig.~\ref{fig11}. We numerically obtain that $t_c(N) \sim N^{0.35}$ (Fig.~\ref{fig12}). Using the convention for the dynamics exponent $z$ defined as $\xi \sim t^{z/2}$ and the relation $N\sim \xi^d$, where $d$ is spatial dimension, we can say that $2/\bar{z}\approx 0.35$, where $\bar{z}=d_uz$ and $d_u$ is the upper critical dimension. This result reveals that the order parameter remains almost unchanged for a long time up to a characteristic time $t_c(N)\approx t_{\infty}$ beyond which it increases rapidly. Thus, we obtain $\xi \sim \Delta r^{-\nu_{\perp}}$ with $\nu_{\perp}=1/2$ and $t_c \sim (\Delta r)^{-\nu_{\parallel}}$ with $\nu_{\parallel}=1$, where $\Delta r=r-r_a$~\cite{absorbing}. Thus, it is obtained that $z/2=\nu_{\perp}=1/2$ and $2/\bar{z}=1/3$, where $\nubar{z}=d_u z$ with $d_u=6$. 

In Ref.~\cite{chung}, it was proposed that a scaling function for the fraction of the nodes in state R of the infinite outbreaks averaged over all configurations is written as $m_t(\Delta r, N, t)=N^{-(\beta+\beta_p)/\nubar_p}m_t(N^{-2/\bar{z}}t, N^{-1/\nubar_p}\Delta r)$ for $\Delta r \equiv r-r_a > 0$. Here $\beta$ is the order parameter exponent, which is zero for the case $\langle k \rangle > 4$. The factor $N^{-\beta_p/\nubar_p}$ is derived using the probability that an infinite outbreak occurs in a given sample. The density of the infectious nodes (denoted as $\rho_{\rm I}$) is obtained as $\rho_{\rm I}(r,N,t)=\partial_t m(r,N,t)$, which becomes, 
\begin{equation}
\rho_{\rm I}(\Delta r, N, t)=N^{-(\beta+\beta_p)/\nubar_p-2/\bar{z}} m(N^{-2/\bar{z}}t, N^{-1/\nubar_p}\Delta r),
\end{equation}
where the exponents $\beta$, $\beta_p$, $\nubar_p$ and $\bar z$ satisfy the following scaling relation.
\begin{equation}
\frac{\beta+\beta_p}{\nubar_p}+\frac{2}{\bar{z}}=1.
\label{exp_z}
\end{equation}
Using $\beta=0$, $\beta_p=1$ and $\nubar_p=3$ for the percolation, one can obtain $2/\bar{z}=1-1/\nubar_p=2/3$. This result is inconsistent with the previous result. 
This implies that the hyperscaling relation for the case $\langle k \rangle > 4 $ does not hold for the MOT.  

\subsection{For $\langle k \rangle=4$}
\begin{figure}[t]
\centering
\includegraphics[width=1.0\linewidth]{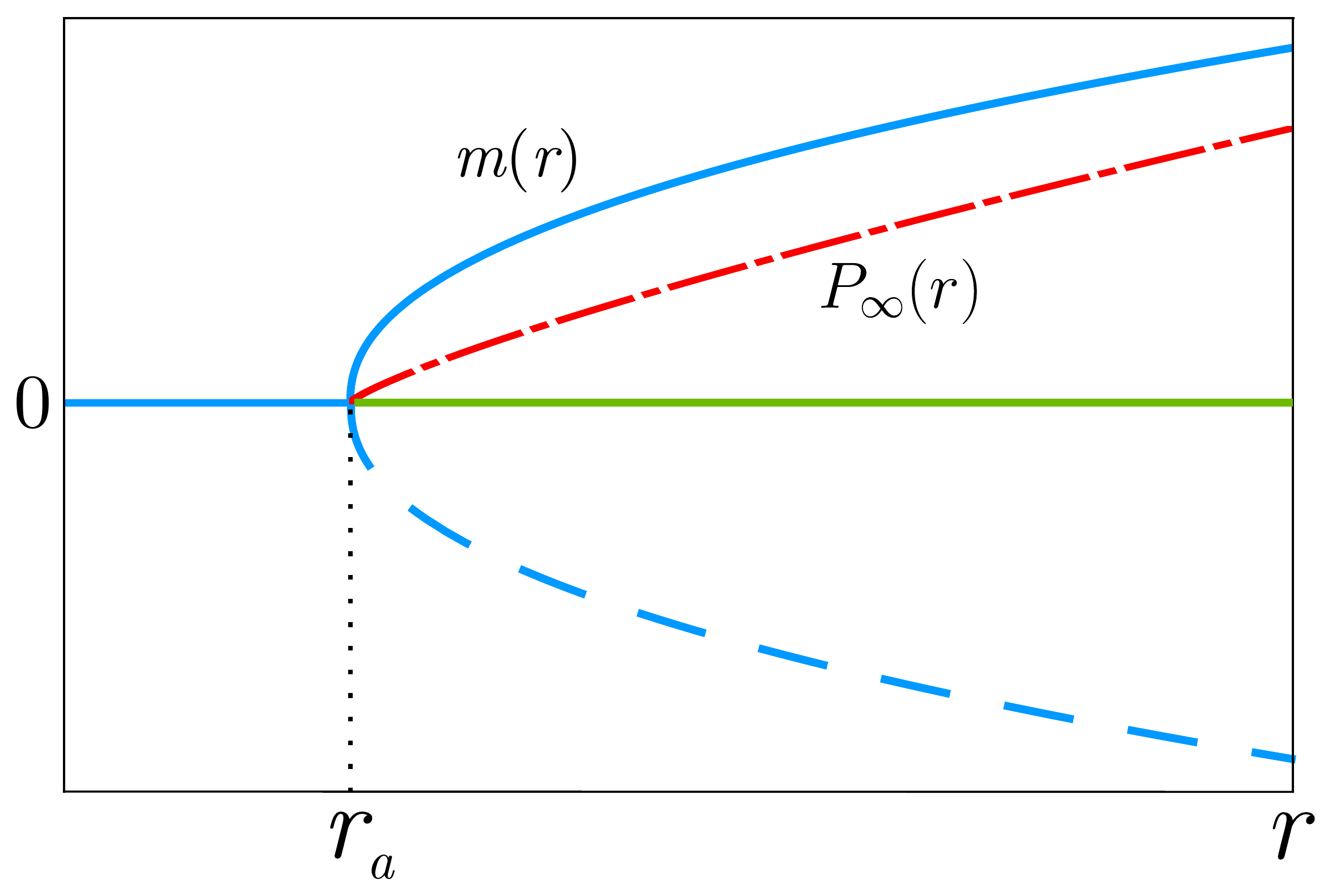}
\caption{Schematic plot of the order parameter $m(r)$ as a function of the reaction probability $r$ for the case $\langle k \rangle=4$. Stable solutions of $m(r)$ are represented by blue (solid) line and curve, whereas unstable solutions are represented by green (solid) line. Physically accessible states are represented by solid lines, whereas inaccessible states are done by dashed lines. The probability $\pinf(r)$ is indicated by a dashed-dotted curve. }
\label{fig13}
\end{figure}

\begin{figure}[h]
\centering
\includegraphics[width=1.0\linewidth]{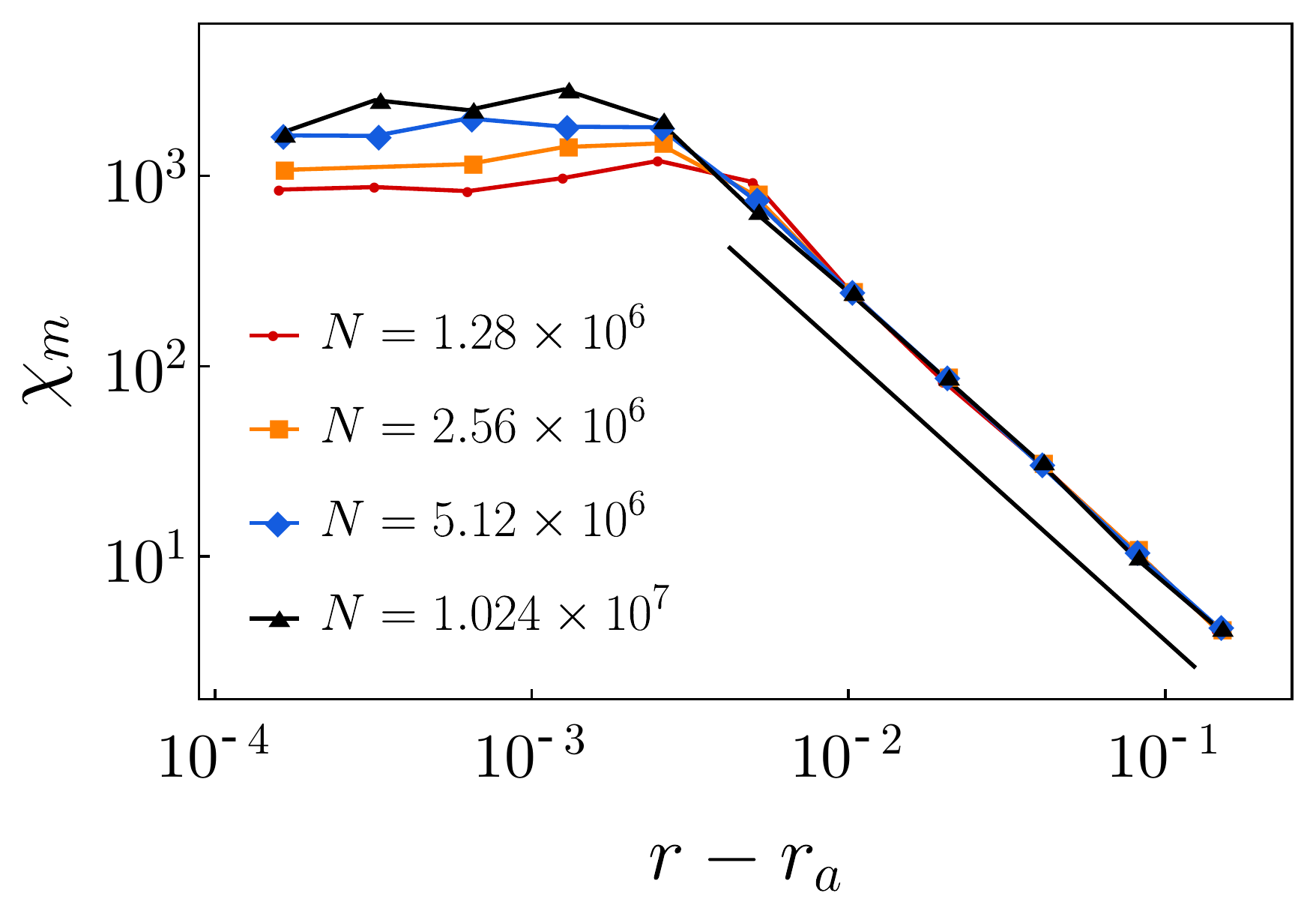}
\caption{ Plot of the susceptibility $\chi_m$ versus the reaction rate for $r >  r_a$ and $\langle k \rangle =4$. Data are obtained from systems of different sizes $N$. Here, the guideline has a slope of $ -1.5 $, which implies that $ \gamma_{a} \approx 1.5$. We remark that data statistics in the plateau region are uncertain because finite and infinite outbreaks are  indistinguishable.}
\label{fig14}
\end{figure}

If $\langle k \rangle=4$, then $r_a=r_b$. Therefore, $a=b=0$ at $r=r_a$, leading to $b^2-4ac=0$. Thus $r_a=r_*$. For this case, a stable solution of $G(m)=0$ for $r < r_a$ is $m=0$. For $r > r_a$, the order parameter behaves as $m(r)=m_u(r) \sim (r-r_a)^{\beta}$ with $\beta\approx 0.5$, so a continuous phase transition occurs (Fig.~\ref{fig13}). In this case, the fluctuation of the order parameter $\chi_m(r)\equiv N(\langle m^2 \rangle - \langle m \rangle^2)$ diverges as $\sim (r-r_*)^{-\gamma_m}$ at $r=r_*$. On the other hand, for the continuous transition, it is not easy to separate the order parameter of size $O(N) $ from that of finite outbreaks of size $o(N)$ near the transition point. Thus, we determine the exponent $\gamma_m$ sufficiently far from the transition point. $\gamma_m$ is measured to be $\gamma_m\approx 1.5$ for $r > r_*$ (Fig.~\ref{fig14}).

\begin{figure}[h]
\centering
\includegraphics[width=1.0\linewidth]{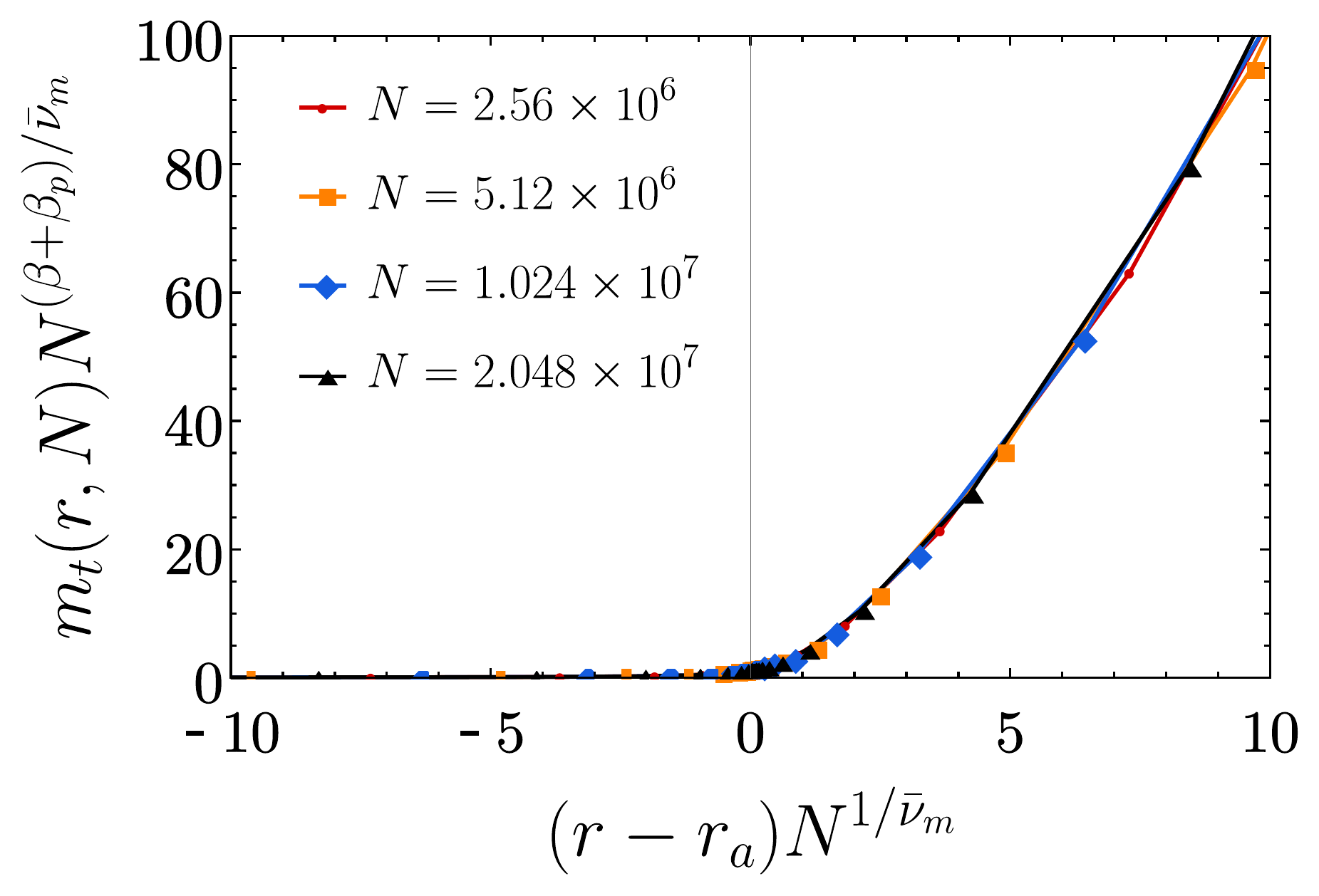}
\caption{Data collapse of the order parameter averaged over all configurations in the form of $m_t(r,N)N^{(\beta+\beta_p)/\nubar_m}$ versus $(r-r_a)N^{1/\nubar_m}$ for $\langle k \rangle=4$. $ \beta= 0.5 $, $ \beta_p =1 $, and $ \bar{\nu}_{m} = 2.5 $ are used. }
\label{fig15}
\end{figure}

In such case, the finite-size scaling method is not useful for determining  the correlation size exponent $\nubar_m$ in the critical region. To determine $\nubar_m$, we used the order parameter defined as $m_t(r) = m(r)P_{\infty}(r)$ averaged over all samples, which is expected to behave as $\sim (r-r_{a})^{\beta + \beta_p}$. We confirm in Fig.~\ref{fig15} that the data from different system sizes are well collapsed onto a single curve with the choice of $\beta$+$\beta_p =1.5$ and $ \bar{\nu}_{m}=2.5$. Thus, $\nubar_m\approx 2.5$ is obtained. Thus, we confirm that the hyperscaling relation $2\beta+\gamma_m=\nubar_m$ holds. 
 
\begin{figure}[h]
\centering
\includegraphics[width=1.0\linewidth]{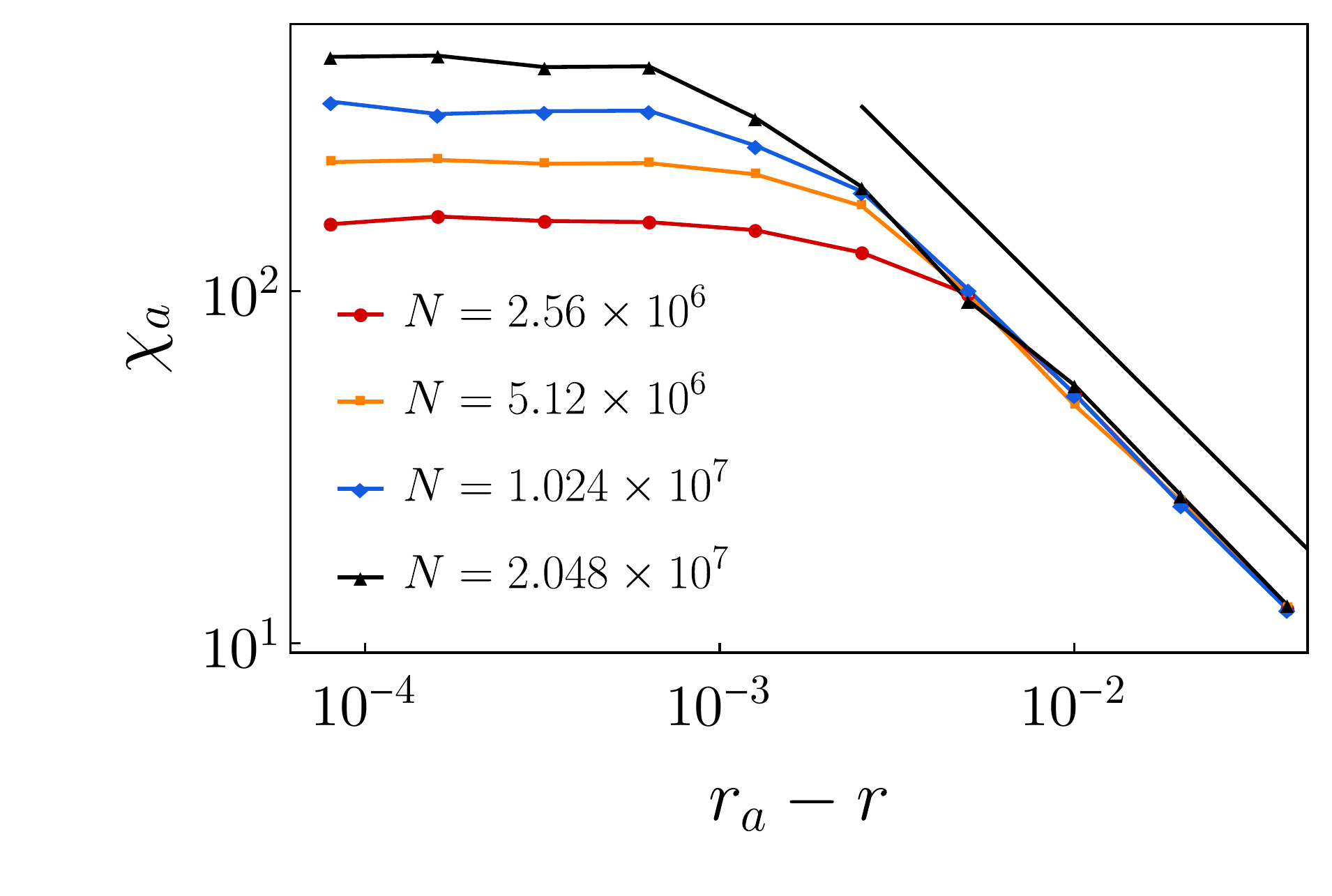}
\caption{Plot of the susceptibility $\chi_a$ versus the reaction rate $\Delta r =r_a-r$ for the case $ \langle k \rangle =4 $. Data are obtained from systems of different sizes $N$. Here, the guide line has a slope of $ -1 $, which implies $\gamma_{a}^{\prime} \approx 1$ for $r< r_a$. We remark that the data statistics in the plateau region are uncertain because finite and infinite outbreaks are indistinguishable.}
\label{fig16}
\end{figure}

\begin{figure}[h]
\centering
\includegraphics[width=1.0\linewidth]{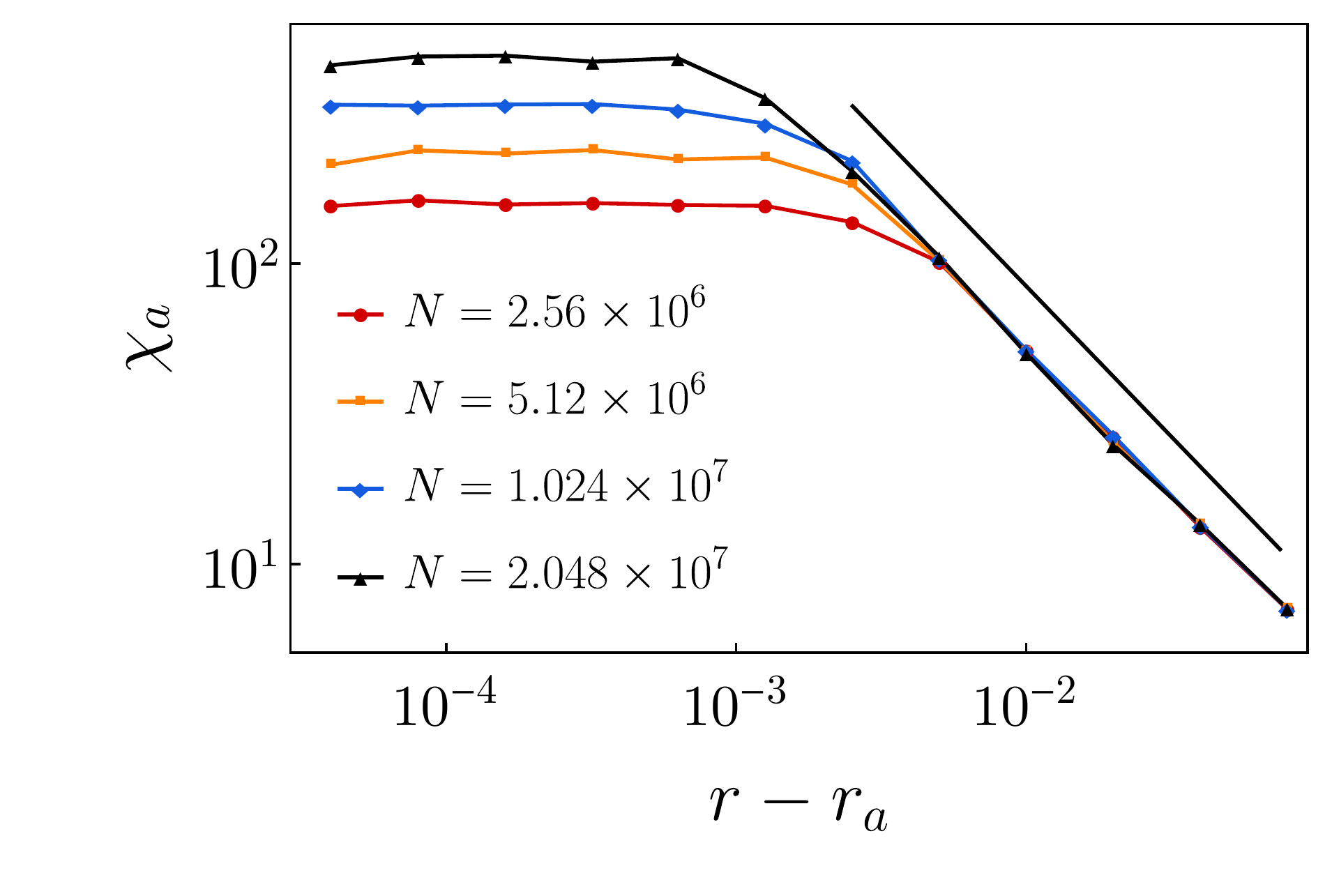}
\caption{ Plot of the susceptibility $\chi_a$ versus the reaction rate $\Delta=r-r_a$ for the case $\langle k \rangle =4$. Data are obtained from systems of different sizes $N$. Here, the guideline has a slope of $-1$, which implies $ \gamma_{a} \approx 1 $. We remark that the data statistics in the plateau region are uncertain because finite and infinite outbreaks are indistinguishable.}
\label{fig17}
\end{figure}

The mean size of finite outbreaks exhibits critical behavior around $r_*$ as $\chi_a=\sum sp_s(r) \sim (r-r_*)^{-\gamma_a}$, where $\gamma_a$ is measured to be $\approx 1$ on both sides of $r_*$ (Fig.~\ref{fig16} and Fig.~\ref{fig17}). 

The point $(r,m)=(r_a,0)$ for $\langle k \rangle=4$ is a tricritical point, because for $\langle k \rangle > (<) 4$, the transition is discontinuous (continuous). See also Ref.~\cite{chung}.

\subsection { For $\langle k \rangle < 4$}
\begin{figure}[h]
\centering
\includegraphics[width=1.0\linewidth]{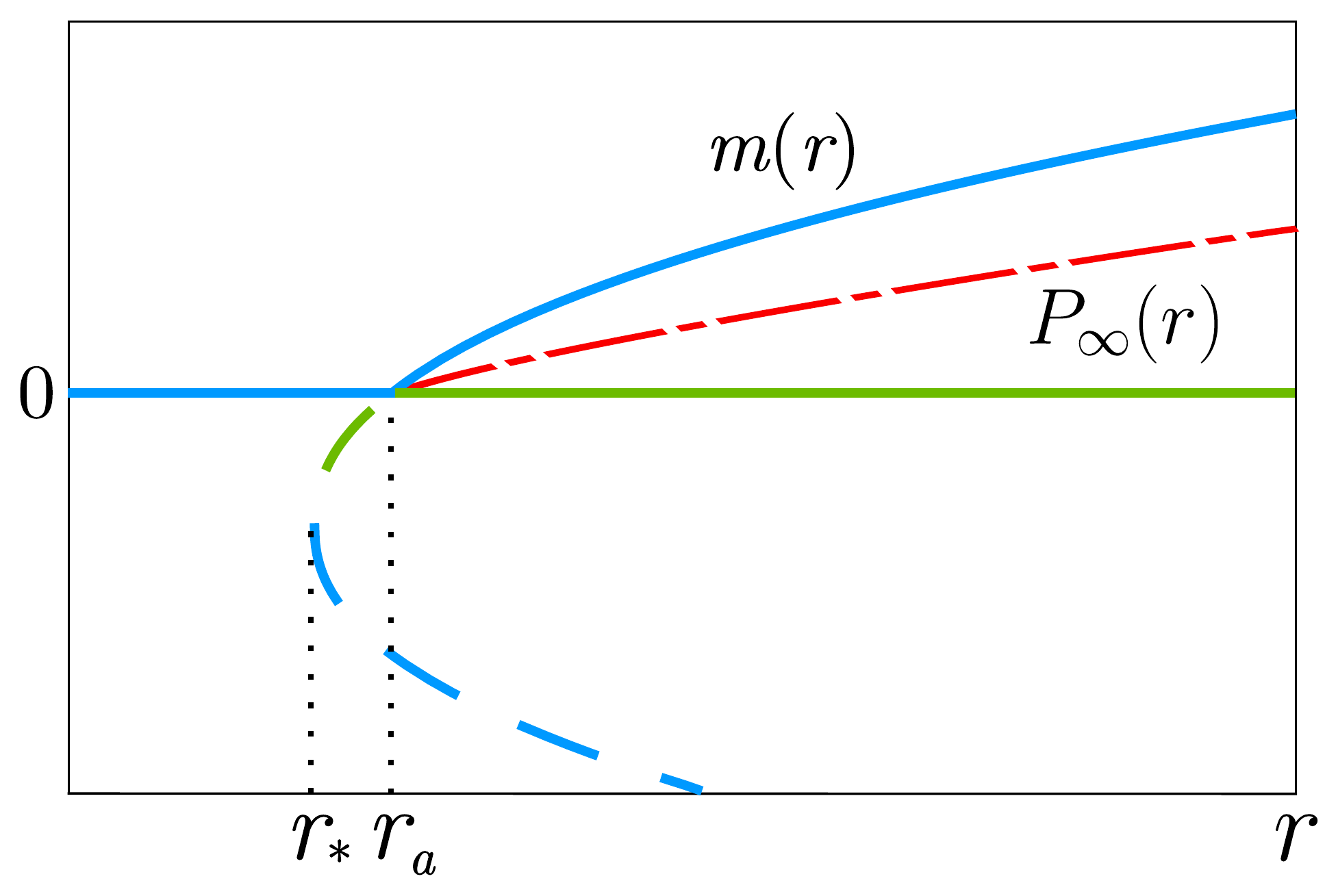}
\caption{Schematic plot of the order parameter $m(r)$ as a function of the reaction rate $r$ for the case $\langle k \rangle < 4$. 
Stable solutions of $m(r)$ are represented by blue (solid or dashed) line and curve, while unstable solutions are done by green (solid or dashed) curve and line. Physically accessible state is represented as solid curve, while inaccessible state is represented as dashed curve. The probability $\pinf(r)$ is represented by dashed-dotted curve.}
\label{fig18}
\end{figure}

\begin{figure}[h]
\centering
\includegraphics[width=1.0\linewidth]{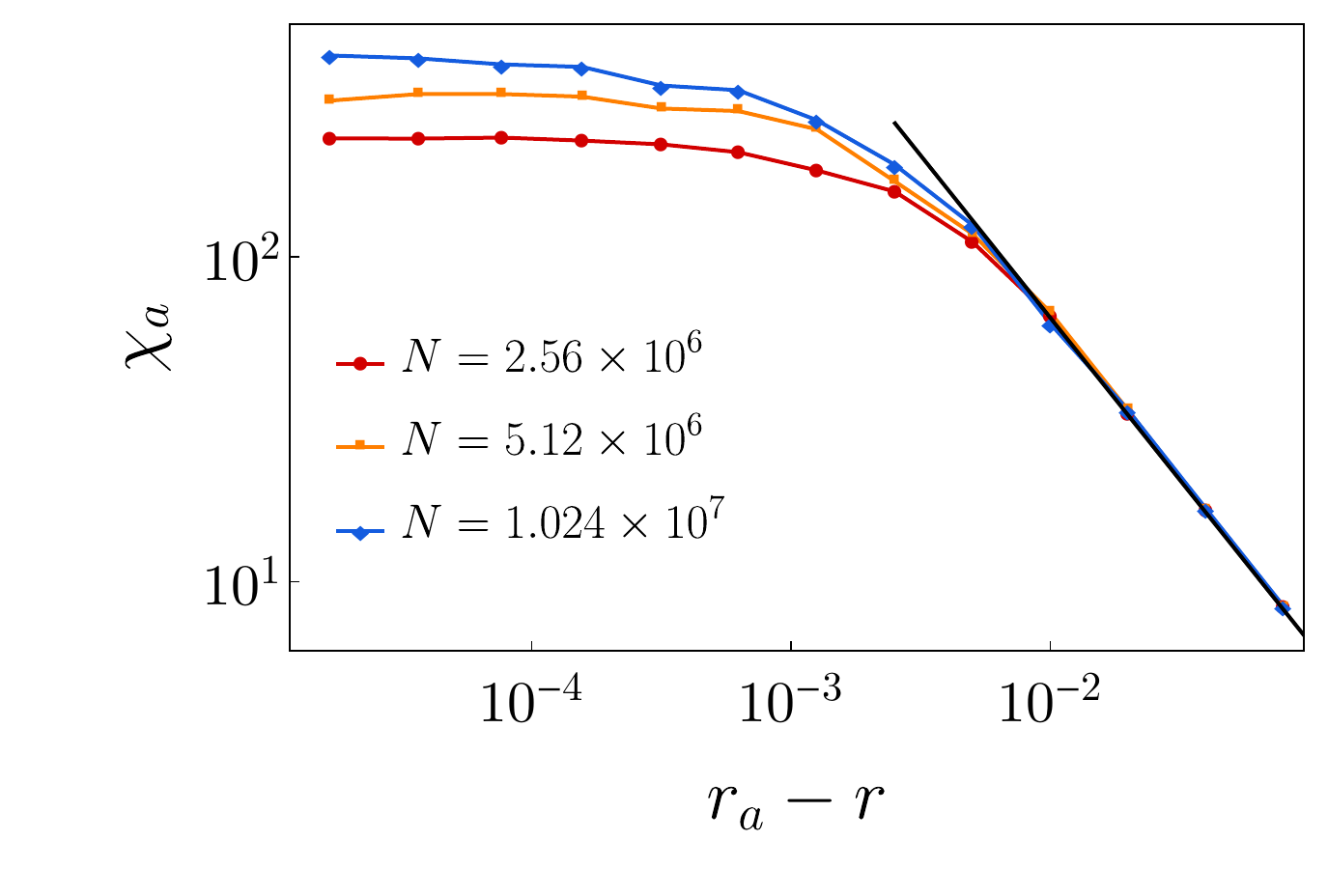}
\caption{Plot of the susceptibility $\chi_a$ versus the reaction rate $\Delta r = r_a-r$ for $r < r_a$. Data are obtained from systems of different sizes $N$. The guideline has a slope of $-1$, implying that the susceptibility exponent $\gamma_a^{\prime}\approx 1.0$. We remark that the data statistics in the plateau region are uncertain because finite and infinite outbreaks are indistinguishable.}
\label{fig19}
\end{figure}

\begin{figure}[h]
\centering
\includegraphics[width=1.0\linewidth]{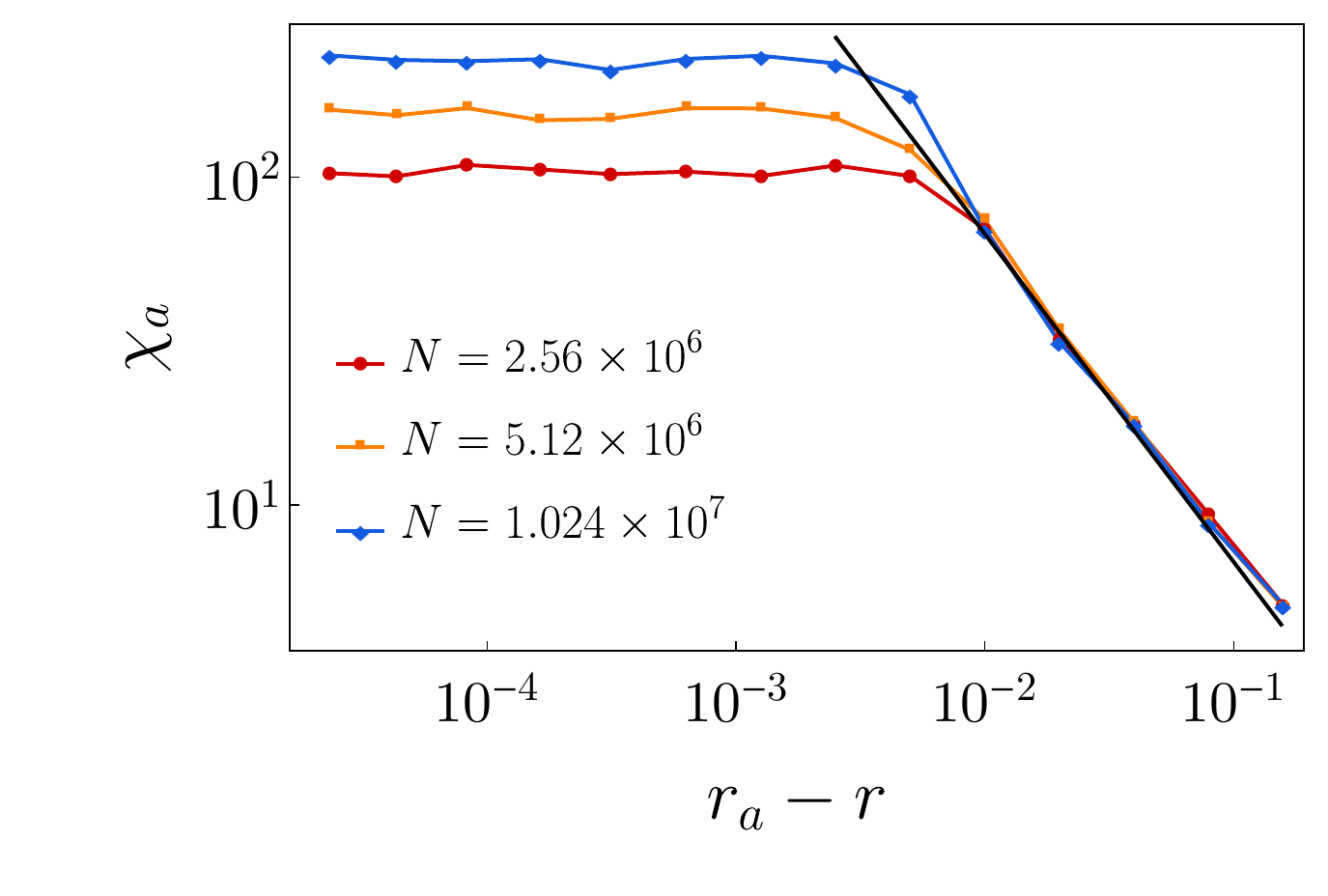}
\caption{Plot of the susceptibility $\chi_a$ versus the reaction rate $\Delta r=r-r_a$ for $ r>r_{a} $. Data are obtained from systems of  different sizes $N$. The guideline has a slope of $-1$. We remark that the data statistics in the plateau region are uncertain because finite and infinite outbreaks are indistinguishable.}
\label{fig20}
\end{figure}

When $\langle k \rangle < 4$, $r_b < r_a$. Further, $r_*$ locates between $[r_b, r_a]$ as shown in Fig.~\ref{fig18}.  At $r=r_*$, $a < 0$, $b > 0$ and $c< 0$, and thus $m_* < 0$. However, for $r > r_a$, the order parameter $m(r)=m_u(r) > 0$, which is physically relevant. The order parameter behaves as $m(r)\sim (r-r_a)$ for $r > r_a$. The fluctuation of the order parameter does not diverge at $r_a$. On the other hand, the probability $\pinf(r)$ behaves as $\pinf(r)\sim (r-r_a)^{\beta_p}$ according to the ordinary percolation theory. The mean size of finite outbreaks $\chi_{a} $ diverges in the critical region around $r_a$ as $ \chi_a \sim (r-r_a)^{-\gamma_a} $, where the exponent is measured to be $\gamma_a \approx 1$ for both $r < r_a$ and $r > r_a$ as shown in Fig.~\ref{fig19} and Fig.~\ref{fig20}, respectively.
 
\section{Summary and Discussion}

We have investigated critical phenomena occurring in a generalized epidemic spreading model, the SWIR model~\cite{janssen} on ER random networks with a controllable mean degree $\langle k \rangle$, particularly when the number of infectious seeds at the beginning is one. The model contains two contagion steps, weakened and infected states. A susceptible node can be either infected or weakened by contacting with an infectious node. The two cases arise stochastically with respective rate. When $\langle k \rangle$ is larger than a characteristic value (depending on the model parameters), a mixed-order transition (MOT) can occur. The nature of this MOT differs from the one of the HPT occurring in $k$-core percolation and the CF model on interdependent network in the following perspective: For the MOT  in the SWIR  model, the order parameter exhibits a discontinuous jump at a transition point without showing any critical behavior. However, other physical quantities such as the mean size of finite outbreaks $\chi_a$ and the probability $\pinf(r)$ that an infinite outbreak occurs in a sample exhibit critical behaviors. Thus, the MOT exhibits the feature of continuous and discontinuous transitions at the same transition point as observed in other systems~\cite{mukamel,dna1}. The critical exponents describing the critical behavior of the SWIR  model belong to the ordinary percolation universality class. For the HPT in $k$-core percolation and in the CF model, the order parameter exhibits a critical behavior following Eq.~(\ref{order}).  
Performing  extensive numerical simulations, we have determined the critical exponents and checked if the conventional scaling relations hold. We found that when a discontinuous transition occurs, a hyperscaling relation does not hold.  

\begin{acknowledgments}
This work was supported by the National Research Foundation of Korea by grant no. NRF-2014R1A3A2069005.
\end{acknowledgments}

\appendix

\hskip 1cm

\section{Classification of phase transitions} 

Here we introduce an analytic method to determine the types of phase transitions as the following cases A-C.  Near a certain point $(r_x, m_x)$, we consider the deviation of the function $G(m(r))$ by $\delta G(m)$ as $r$ and $m$ are perturbed by $\delta r$ and $\delta m$, respectively,  from $(r_x,m_x)$, and set it to zero.
\begin{widetext}
\begin{equation}
\delta G(r_x, m_x) \simeq \dfrac{\partial G}{\partial m}\bigg|_{r_x,m_x}\delta m+\dfrac{\partial G}{\partial r}\bigg|_{r_x,m_x}\delta r+\dfrac{1}{2}\dfrac{\partial^{2}G}{\partial m^{2}}\bigg|_{r_x,m_x}(\delta m)^{2}+\dfrac{1}{2}\dfrac{\partial^{2}G}{\partial r^2}\bigg|_{r_x,m_x}(\delta r)^{2}+\dfrac{1}{2}\dfrac{\partial^{2}G}{\partial r \partial m}\bigg|_{r_x,m_x}(\delta r)(\delta m)+\dots =0
\end{equation}
\end{widetext}

A. For the case $\langle k \rangle > 4$:  at $(r_a, m=0)$, a stable solution exists as $m=0$. Along this line, the derivatives of all orders are zero, and thus any singular behavior does not occur. Thus, divergent behavior does not occur but a discontinuous transition can occur at $r=r_a$.  

B. For the case $\langle k \rangle=4$: at $(r_a, m=0)$, $\dfrac{\partial G}{\partial m}=0$, but $\dfrac{\partial^2 G}{\partial^2 m} <0$ and $\dfrac{\partial G}{\partial r} > 0$. Thus, $(\delta m)^2 \sim \delta r$. The order parameter behaves $m \sim (r-r_a)^{1/2}$. Thus the transition is continuous with the exponent $\beta_m=1/2$.

C. For the case $\langle k \rangle < 4$: at $(r_a, m=0)$, $\delta m \sim \delta r$, so $\beta_m=1$. 

\section{Simulation rule and determination of epidemic spreading time}

During epidemic spreading processes go on, a continuous time variable $t$ passes, which is determined as follows. Suppose that there exists a certain reaction with rate $\alpha$ in the system. Then the probability that the reaction actually occurs between $t$ and $t+dt$ is given as 
\begin{equation}
p_1(t)dt=\alpha (1-\alpha)^{t} dt \approx \alpha e^{-\alpha t}dt.
\end{equation}
In our simulations, once we perform the reaction and regard that the reaction occurs at time $t_1$, which is selected randomly from the probability density function $p_1(t)$. Next, as epidemic spreading proceeds, there exist many possible reactions, e.g., $\ell$ possible reactions with reaction rates $\{ \alpha_{1},...,\alpha_{\ell} \}$, respectively. 
Then  the probability density function $p(t)$ is given as  
\begin{equation}
p(t)=\Big(\sum_{j} \alpha_{j}\Big)e^{-t\sum_{j}\alpha_{j}}.
\label{expDist}
\end{equation}
Then we perform the reaction $j$ with the probability 
\begin{equation}
r_{j}=\dfrac{\alpha_{j}}{\sum_{i=1}^{n} \alpha_{i}}
\end{equation}
and take a time $t_i$ selected randomly from the probability density function (\ref{expDist}). 
We repeat the above process and obtain times $\{t_1, t_2, \dots, t_i, \dots \}$. The final times to reach an absorbing state are given as $t_{\rm finite}=\sum_i t_i$ and $t_{\rm infinite}=\sum_i t_i$ for finite and infinite outbreaks, respectively.

\end{document}